\documentclass[11pt]{article}
\usepackage{latexsym}
\usepackage{amsmath}
\usepackage{amsfonts}
\usepackage{mathrsfs}
\usepackage{dsfont}
\usepackage{enumerate}
\usepackage{verbatim}
\usepackage{hyperref}
\usepackage{tikz}
\usepackage{subfigure}
\usepackage{mathtools}
\usepackage{simplewick}
\usepackage{float}
\usepackage{caption}

 \csname
@addtoreset\endcsname{equation}{section}
\usepackage{amsmath,amssymb}
\usepackage{graphicx}
\usepackage{color}

\begin{document}

\def\be{\begin{equation}}
\def\ee{\end{equation}}
\def\bea{\begin{eqnarray}}
\def\eea{\end{eqnarray}}
\def\nn{\nonumber}

\renewcommand{\thefootnote}{\fnsymbol{footnote}}
\renewcommand*{\thefootnote}{\fnsymbol{footnote}}

\begin{flushright}

\end{flushright}

\vspace{40pt}

\begin{center}

{\Large\sc Topological entanglement entropy and braids in Chern-Simons theory}

\vspace{50pt}

{\sc Hai Siong Tan}

\vspace{15pt}
{\sl\small Division of Physics and Applied Physics,
School of Physical and Mathematical Sciences, \\
Nanyang Technological University,\\
21 Nanyang Link, Singapore 637371}

\vspace{15pt}

{\sl\small
\footnote{On a brief visiting appointment, Summer 2017.}  Enrico Fermi Institute, University of Chicago,\\
Chicago, IL 60637, USA}

\vspace{70pt} {\sc\large Abstract}\end{center}

We explore a web of connections between quantum entanglement and knot theory by examining how topological entanglement entropy probes the braiding data of quasi-particles in Chern-Simons theory, mainly using $SU(2)$ gauge group as our working example.  The problem of determining the Renyi entropy is mapped to computing the expectation value of an auxiliary Wilson loop in $S^3$ for each braid. We study various properties of this auxiliary Wilson loop for some 2-strand and 3-strand braids, and demonstrate how they reflect some geometrical properties of the underlying braids.

\newpage

\tableofcontents

\renewcommand*{\thefootnote}{\arabic{footnote}}
\setcounter{footnote}{0}

\section{Introduction}

In quantum field theory, entanglement entropy is generally an enigmatic non-local quantity that is at present not fully understood. For example, when applied to Yang-Mills theories, the typical definition of bipartite entanglement entropy (being the von Neumann entropy of a reduced density matrix associated with a bipartition of the spatial region into two) does not appear to be compatible with gauge symmetry, which excludes a naive factorization of the Hilbert space, and the notion of entanglement entropy then seems to require some form of refinement \cite{Donnelly}. In certain situations however, we have developed techniques to gain a firmer control. For instance, in string theory, the Ryu-Takayanagi formula \cite{Ryu2} conjectures that the entanglement entropy of certain field theories such as maximally supersymmetric Yang-Mills may be computed via minimal surfaces in its holographic dual. Another known context which admits a clearer understanding is that of topological quantum field theory. It was explained in \cite{Kitaev1} and \cite{Wen1} that for a 2+1D TQFT, the entanglement entropy for a large and simply connected region of linear size $l$ equipped with a smooth boundary would take the form $S_A = \alpha l - \gamma$ where $\alpha$ is a theory-dependent coefficient and $\gamma$ a universal constant known as the topological entanglement entropy. 

Motivated by its significance for characterizing topological orders in fractional quantum hall fluids, the authors of the seminal work in \cite{Fradkin1} set out to compute the entanglement entropy of spatial regions in Chern-Simons theory using surgery methods, including the presence of quasi-particles realized as punctures on the spatial surface. A notable recent work that formulates the results of \cite{Fradkin1} in terms of edge states appeared in \cite{Ryu1, Das} where it was shown that the entanglement entropy detects via an interference effect various topological data of Chern-Simons theory, using the modular tensor category description of the TQFT as the working language. In both these works, the sensitivity of topological entanglement entropy to braiding was essentially studied by using a linear combination of states (amounting to an `interference effect'), with the analysis restricted to two pairs of quasi-particles and pertaining to a few choices of the region that is traced over. As we will elaborate later, if we take the traced region to enclose all the quasi-particles, then the gluing procedure is identical to the usual braid multiplication via concatenation, the end result being identical to the trival unbraided configuration.

We find that there are however other possible choices of the traced region which furnish possibly non-trivial entanglement-related descriptions of the braided configuration. Hence, without alluding to a linear combination of states associated with distinct braids, the entanglement entropy is a non-trivial function of a single braided set of Wilson lines joining quasi-particles. This sets up a simple and physically interesting framework where one could explore relations between quantum entanglement and elements of braid theory.\footnote{See also \cite{Kauffman}, \cite{Brian} and \cite{Vijay} for other lines of exploration based on this theme.}

The modest goal of this paper is thus to begin an exploration of how, after adopting some suitable choices of bipartition, the topological entanglement entropy probes the geometric complexity of braided configurations of quasi-particles and distinguish between different braids. Associated with a braid is the link obtained by its closure which we find to be a useful logical compass in organizing various braid configurations according to their geometrical interpretation as the braid presentations for various links. 

By gluing together punctured discs on distinct copies of three-balls, we map 
the problem of computing the Renyi entropy to calculating the trace of an auxiliary link in $S^3$ via the replica method. The auxiliary link is defined for each choice of bipartition, and this paper is devoted to a study of its geometrical and topological properties. We compute the entanglement
measures (which effectively reduces to computing the Jones polynomial as Chern-Simons VEVs of Wilson loops \cite{Witten}) for a few simple cases: the 2-strand braid with arbitrary number of crossings and the connected sum
of two Hopf links which admits a 3-strand braid representation. The computation essentially reduces to that of the Jones polynomial of the auxiliary link, and relies on us being able to express it as a function of the power index of the density matrix. Thus, we furnish a description of how the entanglement measures distinguish between different braided configurations by being distinct functions of the Chern-Simons level $k$.  Apart from calculating the Chern-Simons VEV, we briefly discuss a couple of topological aspects of the auxiliary link. For the cases considered in this work, we computed the fundamental group of the auxiliary links' exteriors to demonstrate concretely their sensitivity to the braiding parameters and briefly discussed Seifert surfaces associated with them. 

Throughout this work, we will be restricting ourselves to the specific gauge group of $SU(2)$ when computing the entanglement measures. Presumably, it seems that for such a purpose, the complementary edge state approach would be useful, and it should connect our results to conformal field theories. Topological entanglement entropy may potentially be an important quantity as we look forward to developing interferometers for quantum Hall quasi-particles (see for example \cite{Wen2}), and experimental accessibility to topological information such as some knot-theoretic polynomials arising from braiding data would be fascinating. Another area of potential relevance is that of 2+1D quantum gravity where a first step in applying our results would be to generalize them in the context of Chern-Simons theories with $SL(2,\mathbb{R})$ and $SL(2, \mathbb{C})$ gauge groups \cite{Gukov}. In this case, it would be interesting to study how topological entanglement entropy captures the gravitational dynamics associated with the braiding data of the quasi-particles \cite{Carlip, Jackson}.

The outline of our paper goes as follows. We begin by presenting the underlying framework in Section \ref{sec:TopoIntro} where in particular we discuss symmetry properties of the auxiliary link. In Section \ref{sec:Jones}, we compute the Jones polynomial of the auxiliary link for a few cases and thus the Renyi and
entanglement entropies as functions of the Chern-Simons level and the braiding parameter\footnote{By `braiding parameters' we simply refer to integral indices which we can use to characterize the braid word.}
for the 2-strand case. This is done in the context of $SU(2)$ gauge group. In Section \ref{sec:Topology}, we compute the fundamental group of the auxiliary link's exterior and briefly discuss how the genera of their Seifert surfaces scale linearly with respect to the braiding parameters. The paper then concludes with a brief summary and some suggestions for future work.

\section{Topological entanglement of braided quasi-particles on two-sphere}
\label{sec:TopoIntro}

\subsection{General notions and framework}
Our starting point is Chern-Simons theory defined on $S^2 \times (-\infty, 0] \sim B^3$ with the two-sphere
being punctured by $n$-pairs of quasi-particles/holes.  
The particles' trajectories admit a topological classification in terms of a braid description which define
the density matrix associated with the system. In the Lorentzian picture, the 3-ball topology arises from taking time $t \in (-\infty,0]$ and space to be the two-sphere. After pair creation of quasi-particles which carry non-trivial quantum statistics in Chern-Simons theory, one ends up with the topologically equivalent description of a boundary two-sphere punctured with the quasi-particles and connected by braided Wilson lines that extend in the interior of the three-ball. 

\begin{figure}[h]
\centering
\includegraphics[scale=0.7]{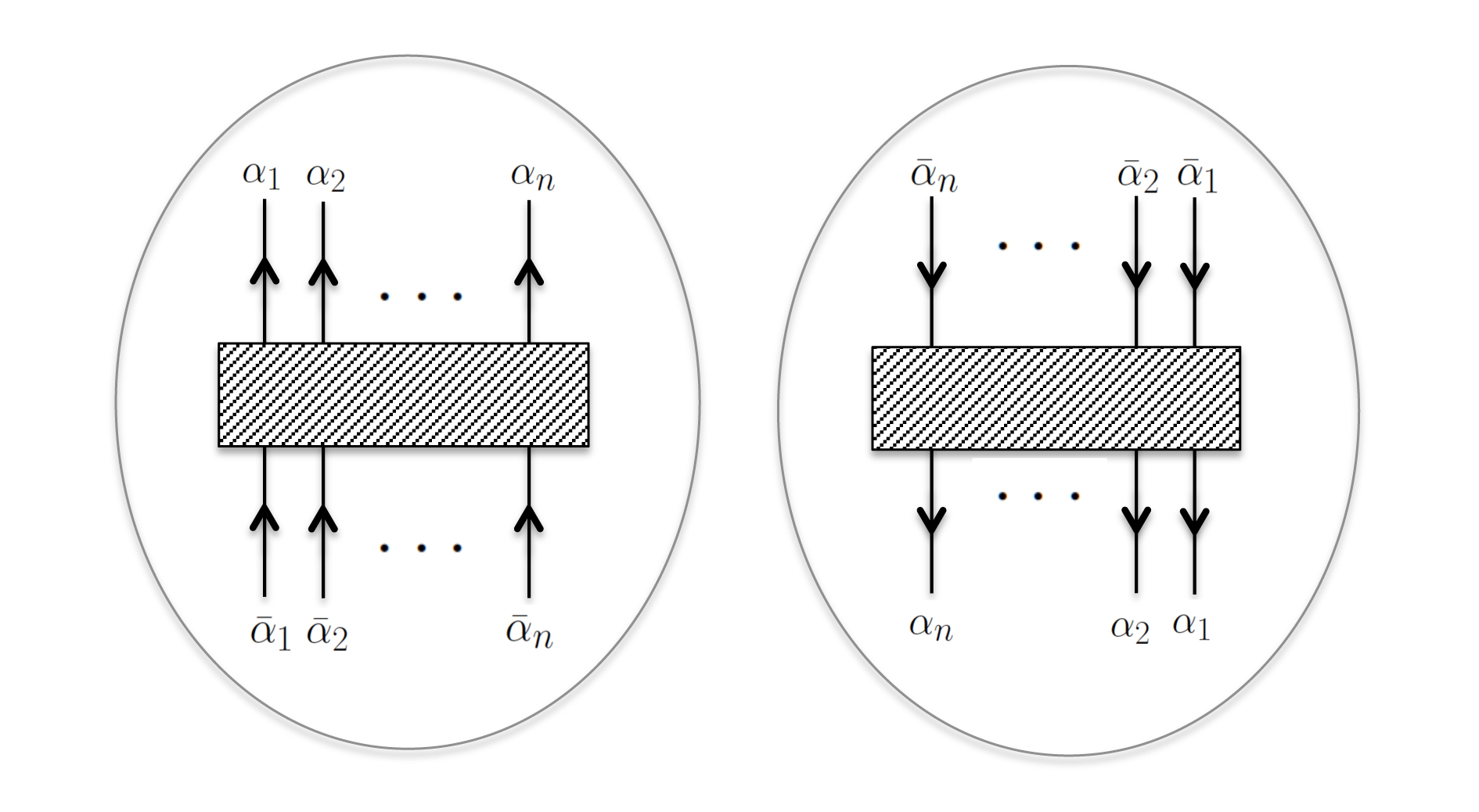}
\caption{We sketch how we associate the state ket with a braided configuration of $n$ pairs of quasi-particles (on the left) and the bra to the dual configuration (on the right). They correspond to the braid words \eqref{braidg1}
and \eqref{braidg2} respectively.}
\label{ketbra}
\end{figure}

As was first explained in Witten's seminal work \cite{Witten}, the path-integral of Chern-Simons theory on the three-ball $B^3$ yields a state $|\varphi \rangle$ living on the boundary $S^2$. Given a braid diagram connecting the quasi-holes on the bottom with the quasi-particles on top (we fix the orientation such that
the Wilson lines flow from the bottom to top in the braid diagram corresponding to the state ket ), one can associate with it a state vector $|\varphi \rangle$ as well as a dual state $\langle \varphi |$ defined by switching the particles with holes, and the nature of the crossings in the braid diagram as sketched in Fig. \ref{ketbra}.

To relate it to a braid description, we can associate $|\varphi \rangle$ and $\langle \varphi |$ with braid words of some length $m$ respectively as
\bea
\label{braidg1}
&&b_{|\varphi \rangle} = \prod^{m}_{i=1} \sigma^{s_i}_{f(i)},\\
\label{braidg2}
&&b_{\langle \varphi |} = \prod^{m}_{i=1} \sigma^{-s_i}_{n-f(i)}.
\eea 
where $s_i \in \mathbb{Z}$, $n$ is the number of pairs of quasi-particles
and $\sigma_k$ refers to the $k^{\text{th}}$ string overcrossing the $(k+1)^{\text{th}}$ string. 
We note that in the standard narrative that relates braids and links, if we close the braids after fixing a braid axis and obtain the link then the link associated with $\langle \varphi |$ has opposite orientation
and is also a mirror image of the original link associated with $|\varphi \rangle$. 

\begin{figure}[h]
\centering
\includegraphics[scale=1]{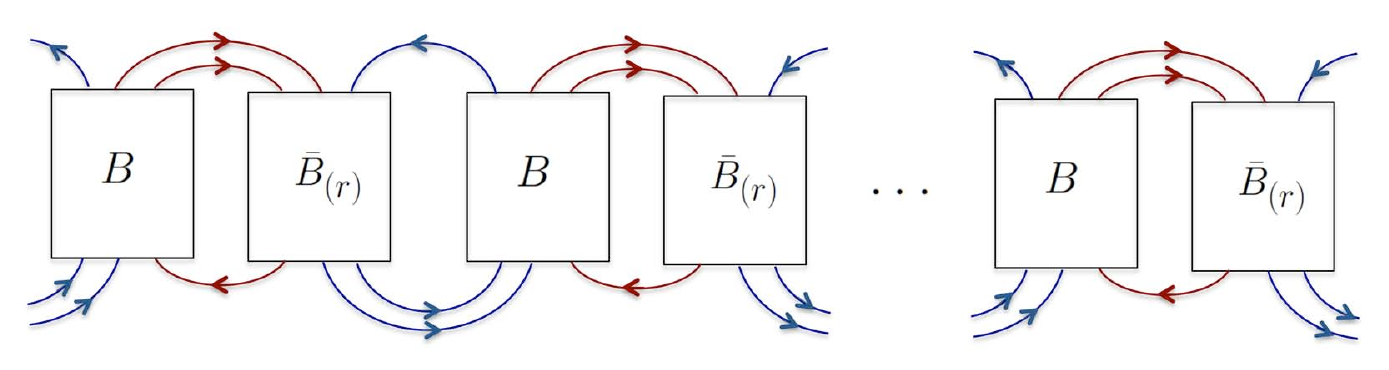}
\caption{Braid diagram showing the auxiliary link $\mathcal{L}_n$ where we have chosen
$R_{\mathcal{B}} = \{ \alpha_2, \alpha_3, \bar{\alpha}_3 \}$. The Wilson lines emerging from either ends 
are identified when we take the trace of $\rho^n$ but are otherwise open punctures in two copies of 
$R_{\mathcal{A}}$. We let $B$ denote a general element of the 3-strand braid group, and $\bar{B}_{(r)}$ denote its mirror image with reversal of orientation. The red lines indicate joining of quasi-particles contained in region $R_\mathcal{B}$ while those in blue depict joining of the particles in $R_\mathcal{A}$. When the trace is taken $\langle \mathcal{L}_n \rangle$ is represented by identifying the corresponding open lines at each end.  }
\label{Generalbraid}
\end{figure}

We are now ready to define the density matrix in terms of an auxiliary link in $S^3$. 
Consider a bipartition of the system into
two subsystems defined by picking a region containing a proper subset of the particles/holes.
Let $R_{\mathcal{B}}$ be a set of discs containing some quasi-particles/holes, then the partial density matrix obtained after tracing over $\mathcal{B}$, i.e. $\rho_A = \text{Tr}_B |\varphi \rangle \langle \varphi |$ is represented by the auxiliary link obtained by joining $n$ copies of this pair of diagram in the following manner: (i) connecting region $\mathcal{B}$ between the diagrams associated with the ket and bra in each factor of\footnote{Henceforth, we would let $\rho$ denote the reduced density matrix $\rho_{\mathcal{A}}$, with $\mathcal{B}$ being the region that is traced over.}  $\rho_{\mathcal{A}}=\rho = |\varphi \rangle \langle \varphi |$, thus perfoming the partial trace (ii) connecting the complementary region $R_{\mathcal{A}}$ between each neighboring pair of $\{ \langle \varphi |, |\varphi \rangle \}$. We are then left with two residual regions $R_{\mathcal{A}}$ in the first ket $|\varphi \rangle$ and last bra
$\langle \varphi |$, as sketched in Fig. \ref{Generalbraid}.

This gluing procedure has an equivalent description of a Chern-Simons path-integral on a 3-manifold that is topologically a 3-ball with punctures. Finally, tracing over the reduced density matrix implies that we identify the remaining boundary areas to obtain an $S^3$ that contains the auxiliary link formed after connecting the punctures in the above fashion. This yields the Renyi entropy after normalizing $\text{Tr} (\rho^n)$
by $(\text{Tr}\, \rho)^n$. By invoking the replica trick, the entanglement entropy follows immediately. 

In this paper, we will let $\mathcal{L}_n$ denote the auxiliary link associated with $\text{Tr}\, (\rho^n )$ and let $\langle \mathcal{L}_n \rangle$ refer to its Kauffman bracket which is, up to a gauge group-dependent factor, the Chern-Simons VEV of the link. Following standard knot theory literature nomenclature, we set the Kauffman bracket of the unknot to be unity. We also
let the Wilson lines to be in the fundamental representation of $SU(2)_k$ Chern-Simons theory, and thus we can write
\be
\label{norm1}
Z \left( \mathcal{L}_n \right) = \text{Tr}\,( \rho^n ) = \langle \mathcal{L}_n \rangle \, S_{0 \frac{1}{2}}
\ee
where $Z \left( \mathcal{L}_n \right)$ is the partition function of Chern-Simons theory with the auxiliary link inserted in $S^3$, $S_{ij} = \sqrt{\frac{2}{k+2}} \sin \left(  \frac{\pi (2i+1)(2j+1)}{k+2}  \right)$ is the modular
S-matrix. It arises in \eqref{norm1} because $Z(\text{unknot}) = S_{0 \frac{1}{2}}$. The Renyi and entanglement entropies then follow to read
\bea
\label{Renyi}
S^{(n)}_R &=& \text{Log}\, S_{0 \frac{1}{2}} + \frac{1}{1-n}\text{Log} \left( \frac{  \langle \mathcal{L}_n  \rangle}{\langle \mathcal{L}_1 \rangle^n} \right), \\
\label{entanglemententropy}
S_{EE} &=& \text{Log}\, ( \langle \mathcal{L}_1 \rangle  S_{0 \frac{1}{2}}) - \frac{1}{\mathcal{L}_1} \partial_n \langle  \mathcal{L}_n \rangle |_{n=1}.
\eea

In the seminal work of \cite{Fradkin1}, it was first explained that topological entanglement entropy in Chern-Simons theory probes the braiding of quasi-particles in an elegant fashion that relates to fusion rules and conformal blocks of the associated WZW conformal field theory living on the spatial two-sphere. This analysis was carried further in \cite{Ryu1} where a connection to edge states and R-matrices (in the language of modular tensor category) was demonstrated. In both papers, the sensitivity to braiding was essentially captured by using a linear combination of states for $|\varphi \rangle$, notably with the discussion restricted to two pairs of quasi-particles and pertaining to a few choices of the regions $R_{\mathcal{B}}$. 

The possibly much richer narrative that we outlined above was however not quite explored previously. 
If we take 
$R_{\mathcal{B}}$ to enclose all the holes (or all the particles), then the gluing procedure is identical to the usual braid multiplication via concatenation, the end result being the standard closure of the braid word
$(b_{|\varphi \rangle} b_{\langle \varphi |})^n$ which by virtue of \eqref{braidg1} and \eqref{braidg2} is nothing but the identity braid.\footnote{Note that when $\mathcal{R}_B$ encloses all quasi-particles, the gluing is represented by concatenating the braid $\langle \varphi |$ with the strand index reversed, i.e.
$k \rightarrow n-k$ in $\sigma_k$ of \eqref{braidg2}. }

This implies that Renyi entropy vanishes for all $n$. There are however other possible choices of $R_{\mathcal{B}}$ which furnish non-vanishing entanglement measures. Hence, without alluding to a linear combination of states associated with distinct braids, the entanglement entropy is generically a non-trivial function of the braid geometry. This furnishes a simple setting where one could relate between the notion of quantum entanglement and elements of braid theory.

For the rest of the paper, we shall embark on an exploration of how, equipped with some suitable choices of the region $R_{\mathcal{B}}$, the entanglement entropies probe the geometric complexity of braided configurations of quasi-particles and distinguish between different braids. Associated with a braid is the link obtained by its closure so it is useful to organize various braid configurations according to their geometrical interpretation as the braid presentations for various links. Nonetheless, as we shall elaborate in detail shortly, the entanglement measures change under Markov moves and we could generically have different entanglement entropies for a pair of braids equivalent up to the two type of Markov moves of stabilization and conjugation.

There is a subtlety that arises for picking a certain $R_{\mathcal{B}}$, that comes with performing the gluing procedure. Let's recall that a standard surgery procedure in $S^3$ involves removing and gluing back a solid torus which is the toroidal neighborhood of some link with a possibly non-trivial homeomorphism of the boundary torus valued in the mapping class group of $T^2$. In our context, we are gluing together a set of punctured discs and identifying punctures. The gluing procedure can be decomposed into two steps - one that translates to taking the partial trace over $R_{\mathcal{B}}$ and the other one corresponding to the field operator product of $\rho$.

It is convenient, as we have illustrated in Fig. \ref{Generalbraid} to order the punctures in a line. Consider the case where some quasi-particles of $R_{\mathcal{B}}$ lie on the right of those in $R_{\mathcal{A}}$ in the diagram corresponding to $\langle \varphi |$ (see Fig. \ref{ketbra}) Now, for every hole in $R_{\mathcal{A}}$, one identifies it with the corresponding particle in the neighboring diagram associated with $|\varphi \rangle$. The gluing is done by connecting the pair with a Wilson line, which then has to cross all the lines connecting the particles in region $R_{\mathcal{B}}$. The choice of either overcrossing or undercrossing is to be made for every Wilson line crossed. Thus, in general, the gluing procedure that identifies the copies of $R_{\mathcal{A}}$'s in each neighboring pair of ket and bra is accompanied with an indication of the choice of crossings among the Wilson lines whenever such ambiguity arises. 

We could keep track of any choice by indicating a braid word for each quasi-particle in any subregion of $R_\mathcal{B}$ that lies on the right of those in $R_\mathcal{A}$ for $\langle \varphi |$. But for the rest of the paper, we will not discuss this subtlety further. Our working examples involve choices of the region $R_{\mathcal{B}}$ such that the auxiliary link can be constructed uniquely once we indicate the quasi-particles/holes that $R_{\mathcal{B}}$ contains.

\subsection{Amphichirality and invertibility of the auxiliary link}

Prior to presenting specific examples, it is useful to first recognize some salient universal symmetry properties of the auxiliary link constructed from the gluing procedure described above. 

The auxiliary link turns out to be one that is always fully amphichiral. We first demonstrate that it is always invertible. Define an orthogonal set of Cartesian axes such that the link projection lies in the $zy$-plane. 
A rotation of $\pi$ about the $z$-axis turns the braid diagram into its reversal as depicted in Fig. \ref{symmetry} below. This is consistent with the fact that our choice of Wilson lines being directed from  quasi-holes to quasi-particles is merely a convention that should not matter to any measures of quantum entanglement and thus the auxiliary link. 

A similar argument can be run to demonstrate it is amphichiral. To see that it is ambient isotopic to its mirror image, perform a rotation of $\pi$ about the $x$-axis, and then about the $y$-axis, after which it is easy to see that we end up with the braid diagram with all the overcrossings being replaced by undercrossings and vice-versa, with the orientation preserved. Thus, the auxiliary link is fully amphichiral by virtue of the symmetry induced from the braids diagrams associated to the ket and bra. We sketch these observations in Fig. \ref{symmetry}.

\begin{figure}[h]
\centering
\includegraphics[scale=1.0]{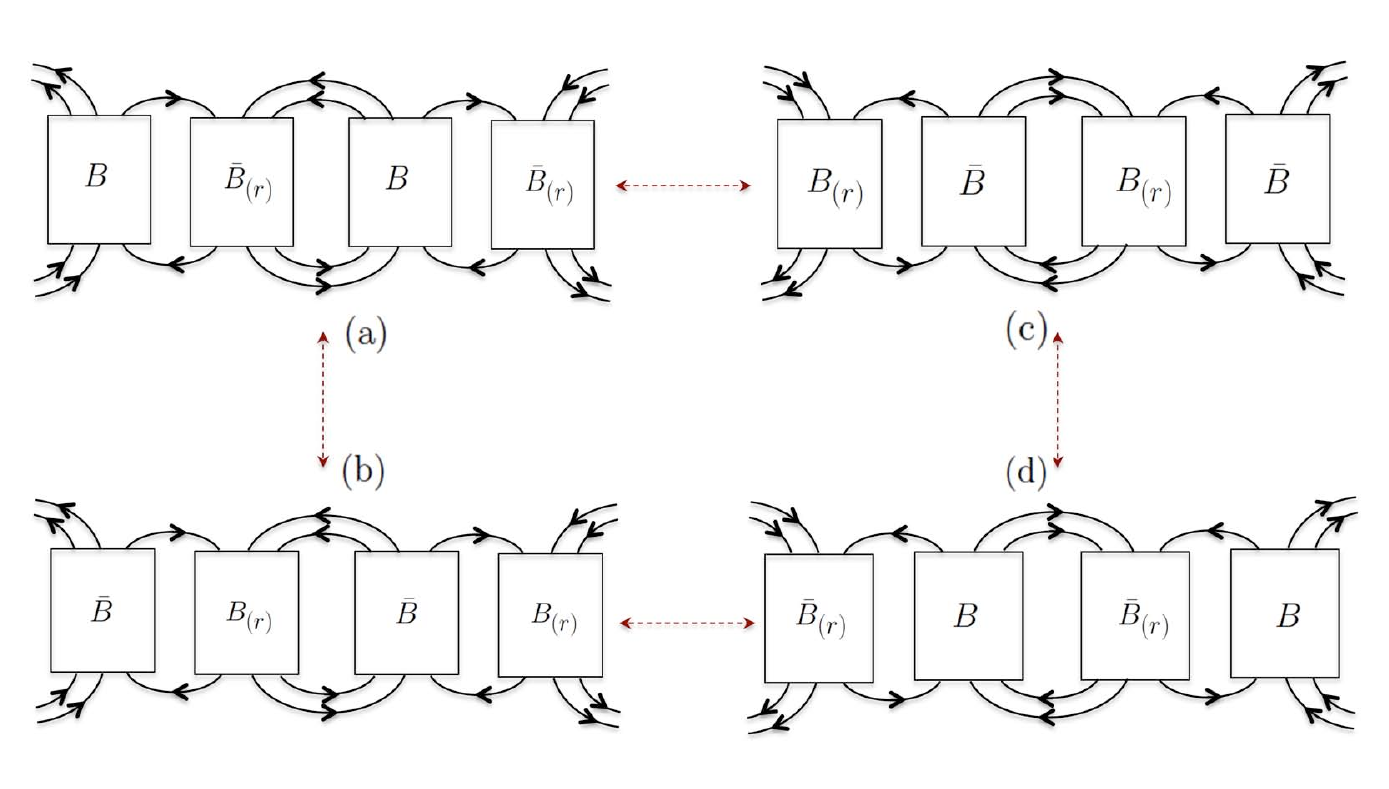}
\caption{A series of diagrams revealing a couple of $\mathbb{Z}_2$ symmetry properties of the auxiliary link. A rotation about the vertical axis takes (a) to (c) which is its reversal whereas another rotation takes (a) to (b) which is its mirror image. Thus, the auxiliary link is fully amphichiral. In the above, we have taken $R_{\mathcal{B}} = \{\alpha_3, \bar{\alpha}_3 \}$ without loss of generality. }
\label{symmetry}
\end{figure}

This property ensures the reality of the Renyi entropy since the Jones polynomial has to be palindromic. It also implies that the Renyi and entanglement entropies cannot distinguish a braid configuration with a dual one in which all overcrossings are replaced by undercrossings. Closure of this pair yields the link and its mirror image, and thus in this sense, we can say that the entanglement entropies cannot detect chirality.


\subsection{Entropy as a function of braid words and knots}
Given a specific braid that defines the density matrix, we wish to compute its associated quantum entanglement entropy measures. There is a distinguished link associated with the braid which is
the link obtained via its closure. And thus it is natural to ask how is the geometric complexity of 
the link captured by the quantum entanglement entropies for various choices of bipartition?

We first note that the entanglement measures (that comes together with a choice of $R_\mathcal{B}$) could
be different for two braids of which closure yield the same link. They do not respect the more restrictive notion of braid equivalence but in a similar spirit, one could readily write down the set of choices of $R_\mathcal{B}$ for two braid configurations related by Markov moves and which also yield identical entanglement entropies. 

Recall that Markov's theorem asserts that any two braid closures related by two types of Markov moves yield the same knot/link (see for example \cite{Lick} and \cite{Birman}). The two moves are 
\begin{enumerate}
\item Changing an element of the braid group $\mathcal{B}_n$ to a conjugate element
in that group, 
\item Changing the element $b$ to $i_n ( b ) \sigma_n^{\pm 1} \in \mathcal{B}_{n+1}$, where $i_n : \mathcal{B}_n \rightarrow \mathcal{B}_{n+1}$ is the inclusion that disregards the $(n+1)^{\text{th}}$ string. 
\end{enumerate}

We have seen that the replica trick maps the original knot/link associated with the braid (that defines $|\varphi \rangle$) to an auxiliary knot/link, yet it is quite easy to see that the latter's closure is not generally invariant under the two Markov moves for generic choices of $R_\mathcal{B}$. Corresponding to
two braid configurations which are Markov equivalent, one could nonetheless specify the corresponding choices of $R_\mathcal{B}$ such that the entanglement measures are equivalent. 

Let the pair of regions be denoted by $R_\mathcal{B}, R^{(M)}_\mathcal{B}$ respectively. 
For Markov move (i), if each quasi-particle contained in $R^{(M)}_\mathcal{B}$
is connected to its corresponding one in $R_\mathcal{B}$ by a Wilson line (i.e. we let the same conjugation element act on the quasi-particles in $R_\mathcal{B}$), the entanglement
measures will be identical.
The second Markov move (ii) or stabilization relates the braid to another with one more strand with the braid generator $\sigma^{\pm 1}_n$ multiplied to the original word. The closure of the auxiliary braid is invariant up to some multiplicative factors of the VEV of the unknot if we replace 
$$
\bar{\alpha}_n \rightarrow \bar{\alpha}_{n+1}
$$
for its inclusion in $R^{(M)}_\mathcal{B}$ in the braid. Further, if we let both $\alpha_{n+1}$ and $\bar{\alpha}_n$ be included or excluded in $\mathcal{R}_B$ then the entanglement measures remain unchanged as shown in Fig. \ref{markov1} below.  For the remaining cases as depicted in Fig. \ref{markov2}, the Renyi and entanglement entropies will gain an additional factor of $\text{Log}\, \langle \bigcirc \rangle$, where $\langle \bigcirc \rangle$ is the VEV of the unknot. Apart from these caveats, we find it useful to compare how the entanglement measures correlate with the geometric nature of the knot/link obtained via the closure of the braid associated with $|\varphi \rangle$.

\begin{figure}
\centering
\includegraphics[scale=1]{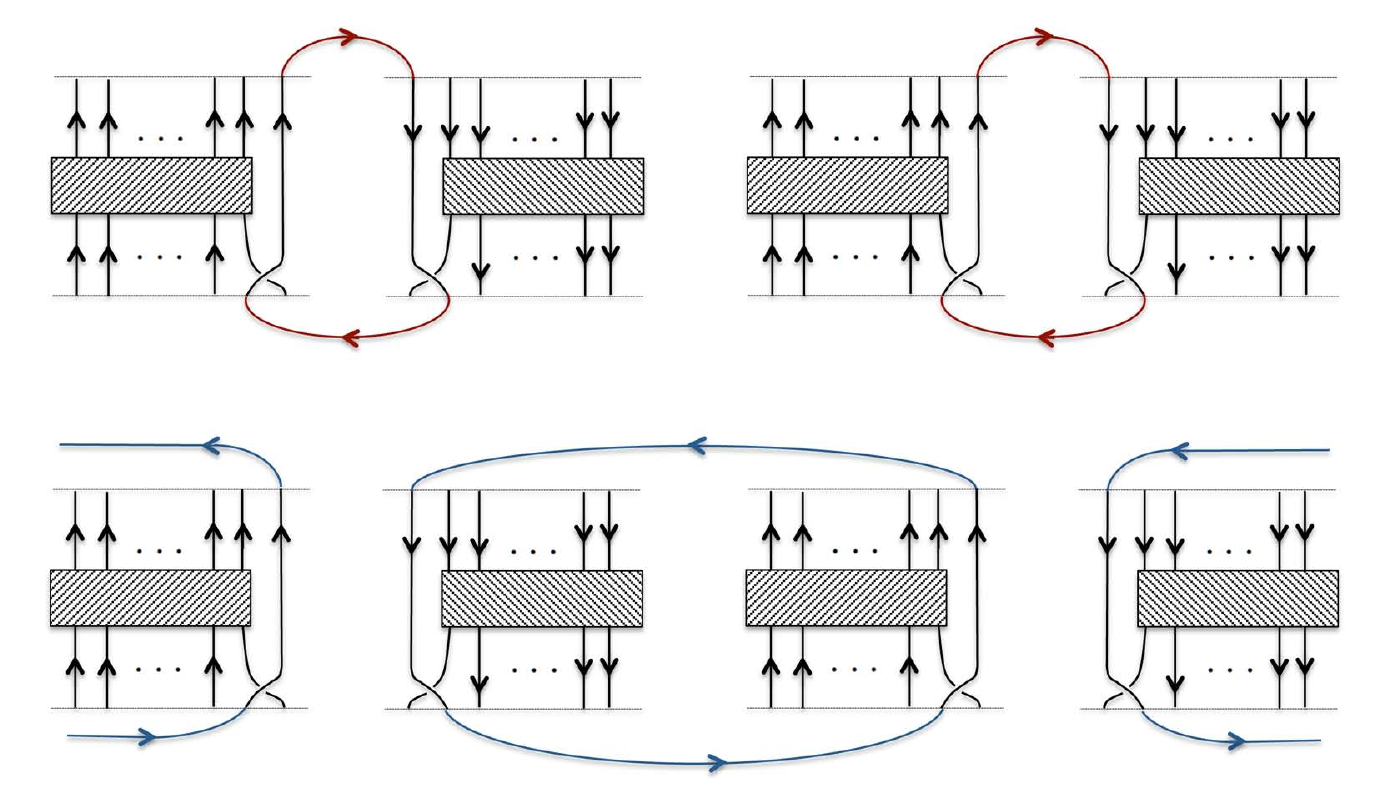}
\caption{The diagrams in the upper and bottom rows depict the cases where $R_{\mathcal{B}}$ includes or excludes both $\alpha_{n+1}, \bar{\alpha}_n$ respectively. For every factor of $\rho$, the trace of $\rho^n$ gains a factor of the unknot's VEV. In computing the Renyi entropy, this is cancelled by the normalizing factor of $(\text{Tr} \, \rho)^n$ and hence they are equivalent to the diagrams prior to the Markov move of stabilization. }
\label{markov1}
\end{figure}

\begin{figure}
\centering
\includegraphics[scale=1]{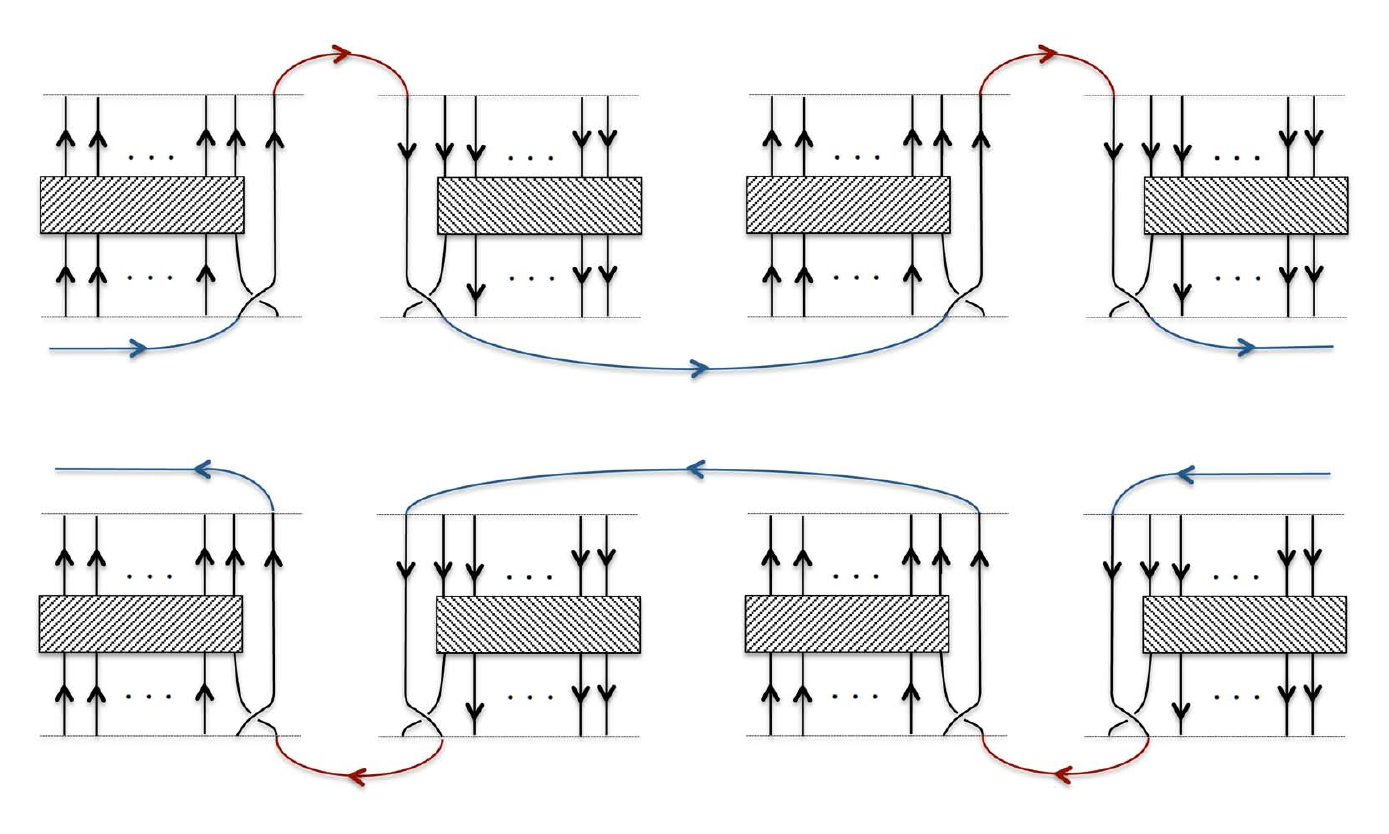}
\caption{The diagrams in the upper and bottom rows depict the cases where $R_{\mathcal{B}}$ includes either $\alpha_{n+1}$ or $\bar{\alpha}_n$ respectively. The trace of $\rho^n$ gains a factor of the unknot's VEV independent of $n$.  }
\label{markov2}
\end{figure}


\subsection{On braid data and the choice of $R_{\mathcal{B}}$}

The general problem of determining the Renyi entropy $S^{(n)}_R$
for a braided configuration is typically difficult
as it may not always be straightforward to determine the Jones polynomial as an explicit function in $n$. As
we shall see shortly, even in the simple examples that we consider in this work, the derivation
can be quite involved. In the following, we shall first make some general remarks on $Z(\mathcal{L}_n)$ with regards how it depends on the braid data and choice of $R_{\mathcal{B}}$ before plunging into a full computation of the entropies in the subsequent section. 

\begin{figure}[h]
\centering
\includegraphics[scale=1.1]{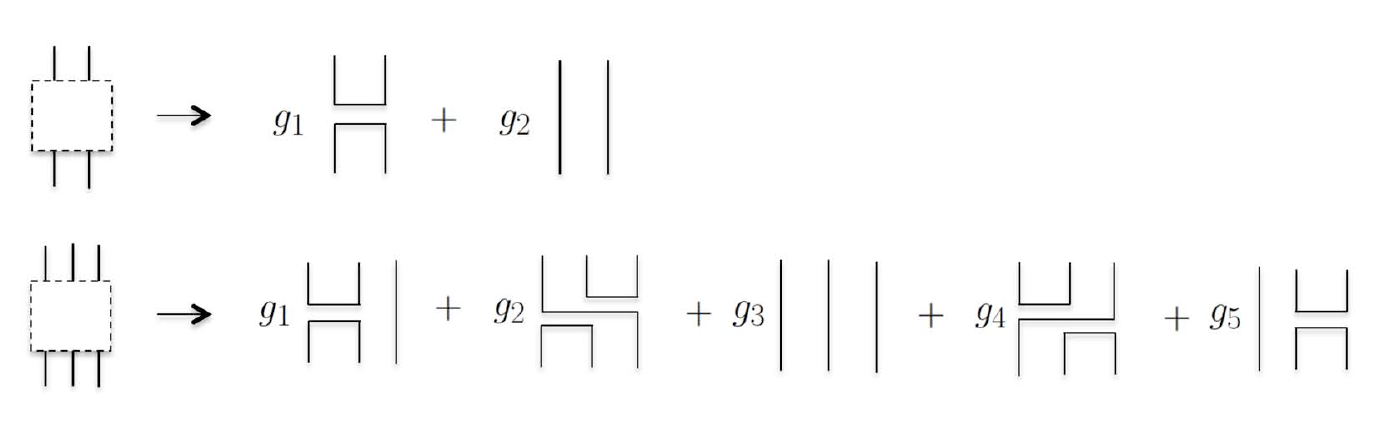}
\caption{The skein relations can be used to resolve the crossings in an arbitrary braid such that we end up with a fixed  set of non-intersecting Wilson lines as portrayed for the case of 2-strand and 3-strand above. The coefficients $g_i$ capture the braiding data. }
\label{resolve}
\end{figure}

For every $m$-strand braid, the end of resolving all the crossings is 
described by a set of braid diagrams with punctures connected by
non-intersecting lines. Powers of the density matrix and its trace can always
be expressed in such a basis. The cases of 2-strand and 3-strand braids are
sketched in Fig. \ref{resolve}, we denote all the factors arising from the skein relations to be captured by variables $g_i$ multiplied to diagrams of non-intersecting Wilson lines at the end of
the resolving tree. They are denoted by $\mathcal{D}_i$ in the following discussion. 

Let's consider the 3-strand braid for definiteness. Expanding
the state ket, we can write
\be
|\varphi \rangle = \sum^5_{i=1} g_i \mathcal{D}_i, \qquad
\langle \varphi | = \sum^5_{i=1} \bar{g}_i \mathcal{D}_{6-i}, \qquad
\rho = | \varphi \rangle \langle \varphi | = \sum_{i,j} g_i \bar{g}_j 
\left( \mathcal{D}_i 
\bigoplus_{\mathcal{B}} \mathcal{D}_{6-j} \right),
\ee
where $\bar{g}_i$ denotes the complex conjugate of $g_i$ and we have abused the direct sum symbol to denote connecting the braid diagrams with some specific choice of $R_{\mathcal{B}}$. We can then write down powers of $\rho$ in terms of these variables, for example,
\be
\rho^2= \sum_{i,j,k,l} 
g_i \bar{g}_j g_k \bar{g}_l \left( \mathcal{D}_i \bigoplus_{\mathcal{B}} \mathcal{D}_{6-j} \right) 
\bigoplus_{\mathcal{A}} \left(  \mathcal{D}_k \bigoplus_{\mathcal{B}} \mathcal{D}_{6-l} \right) 
\ee

The Renyi entropy can thus be expressed as a linear combination of terms, each of which is a product
of braid data and a union of Wilson loops that is a function only of the number of quasi-particles and choice of $R_{\mathcal{B}}$. We can write suggestively
\be
\text{Tr}\, ( \rho^n )= \sum_{i_1,\ldots,i_n, j_1, \ldots, j_n} \prod^n_{k=1} g_{i_k} \bar{g}_{j_k}
\bigoplus_{\mathcal{A}} \langle Y_{i_k \bar{j}_k} \rangle,\,\,\, 
\ee
where $Y_{i_k \bar{j}_k} \equiv \left( \mathcal{D}_i \bigoplus_{\mathcal{B}} \mathcal{D}_{6-j} \right)$
and $\langle Y_{i_k \bar{j}_k} \rangle$ denotes the Chern-Simons VEV which is obtained after identifying the open ends of $Y_{i_k \bar{j}_k} $. 
In this form, we see that the braid data is contained in the factors of $g_m$ and its conjugate whereas the dependence on $R_{\mathcal{B}}$ is expressed through the $Y_{i\bar{j}}$'s.


\subsection{Universal cofficient constraints from braid concatenation}
\label{universal}

\begin{figure}[h]
\centering
\includegraphics[scale=0.55]{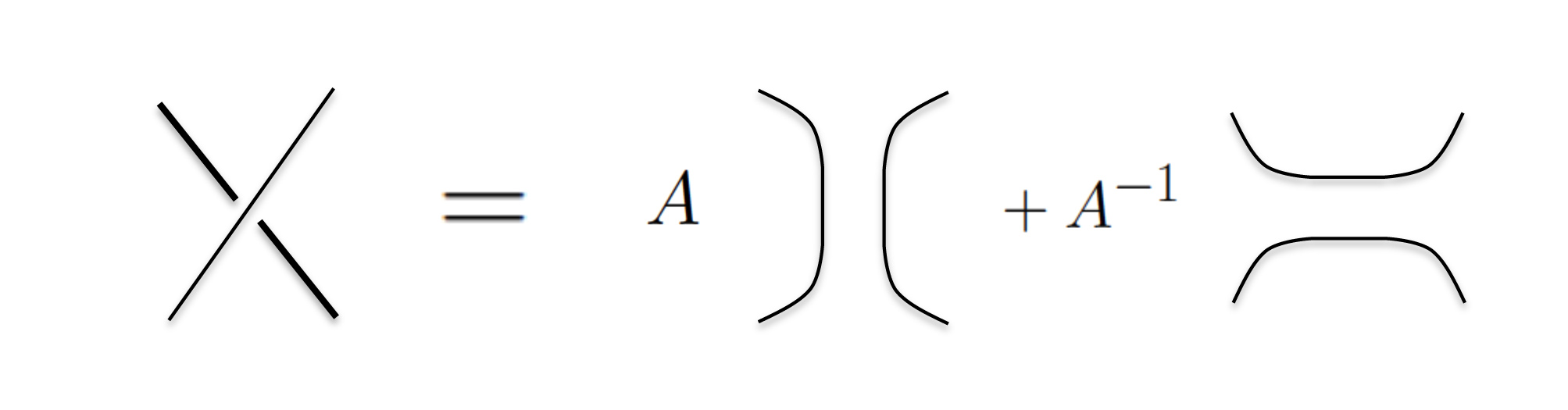}
\caption{This skein relation together with $\langle L \cup \bigcirc \rangle = (-A^2 - A^{-2}) \langle L \rangle$ for any link $L$ are those for the bracket or Jones polynomial in variable $A$. We also take $\langle \bigcirc \rangle =1$.  }
\label{basicskein0}
\end{figure}

The coefficients $g_i$ are functions of the braid word. In the 2-strand case, it is simple to derive them from the skein relations. If we adopt $R_{\mathcal{B}}$ to be the entire set of quasi-particles or holes, then the density matrix is represented by the concatenation of the braid words and is always equivalent to the trivial braid. We can use this fact to obtain useful constraints for the coefficients $g_i$ which are `universal' in the sense of being independent of the choice of $R_{\mathcal{B}}$. In the following, we work in the context of $SU(2)$ Chern-Simons theory where the skein relations is depicted in Fig. \ref{basicskein0}.

\begin{figure}[h]
\centering
\includegraphics[scale=0.3]{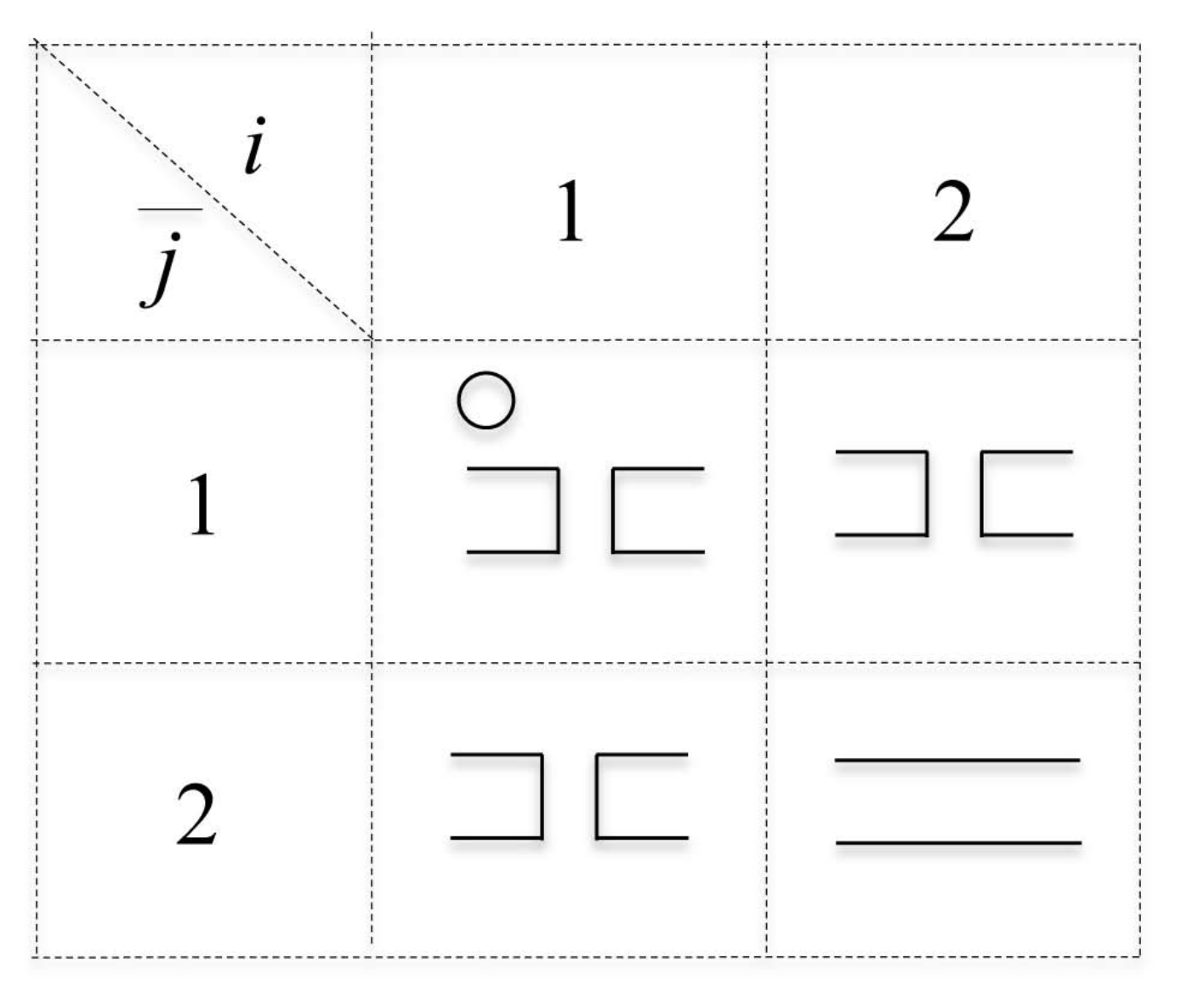}
\caption{This table displays the configurations of non-intersecting Wilson lines $Y_{i\bar{j}}$ for the 2-strand case, with $R_{\mathcal{B}} = \{ \alpha_1, \alpha_2 \}$. The $(1,1)$ entry depicts the split union of an unknot with two open Wilson lines.  }
\label{Tab1}
\end{figure}

First, let us consider the two-strand case, where as depicted in Fig. \ref{resolve}, we merely have two coefficients $g_1, g_2$. The braid is defined solely by an integer counting the number of over/under-crossings. Taking $R_{\mathcal{B}} = \{ \alpha_1, \alpha_2 \}$, we display the various $Y_{i\bar{j}}$ in Fig. \ref{Tab1}. Since from braid concatenation, it is clear that $\mathcal{L}_1$ is an unbraided 2-strand, we obtain the coefficients to satisfy
\be
\label{constraint1}
|g_2|^2 = 1, \qquad
h_0 |g_1|^2 + g_2 \bar{g}_1 + \bar{g}_2 g_1 = 0.
\ee
where $h_0= -A^2-A^{-2}$. In this case, we can invoke the skein relation and the form of the general braid word which is just $\sigma^{\beta}, \beta \in \mathbb{Z}$ to compute
$g_{1,2} = g_{1,2} (A)$ easily. We find them to be
\be
\label{2strandg}
g_2 = A^\beta, \qquad
g_1 = \frac{1}{h_0}\left( -A^\beta + (-A^{-3} )^\beta \right)
\ee
which indeed satisfy \eqref{constraint1}. To derive this 
relation, we note that by the skein relation, every crossing yields two different arc diagrams and
 the only combination to obtain two unbraided lines arises from taking the first arc diagram for every crossing and thus we have a factor of $g_2 = A^\beta$. For every $n$ factor of the second arc diagram, we have an additional 
factor of $n-1$ unknots together with a factor of $A^{-1}$. As the order does not matter, we can use a binomial expansion to keep track of the terms and thus obtain
\be
\label{usefulresult}
g_1 = \frac{1}{h_0} A^\beta \sum_{i=1}^n  {}^n C_i \left(  \frac{A^{-1}h_0}{A} \right)^i 
=\frac{1}{h_0}A^\beta \left[ (1+ \frac{A^{-1}h_0}{A} )^\beta - 1 \right] 
= \frac{1}{h_0} ( -A^\beta + (-A^{-3})^\beta )
\ee

\begin{figure}[h]
\centering
\includegraphics[scale=0.8]{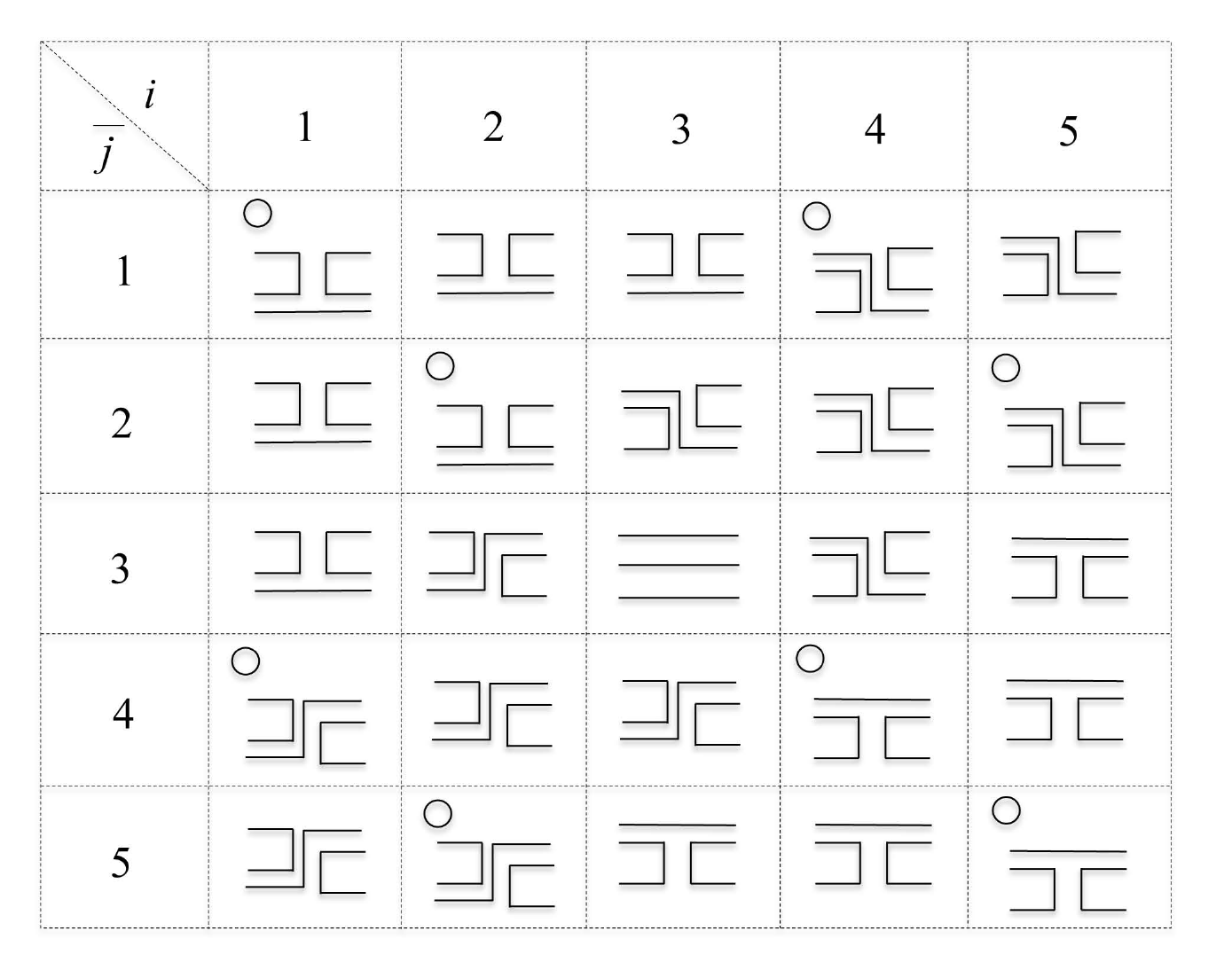}
\caption{This table displays the configurations of non-intersecting Wilson lines $Y_{i\bar{j}}$ for the 3-strand case, with $R_{\mathcal{B}} = \{ \alpha_1, \alpha_2 ,\alpha_3\}$. They lead to the constraints on coefficients $g_i$ in eqns. \eqref{c1}-\eqref{c4}. }
\label{tab}
\end{figure}

As another example, let us consider the 3-strand braid where we now have five coefficients $g_i, 1 \leq i \leq 5$ as mentioned previously, with 
the various possible $Y_{i\bar{j}}$ sketched in Fig. \ref{tab}. Again purposefully picking $R_{\mathcal{B}}$ to be the entire set of quasi-particles and asserting that the sum is an unbraided 3-strand, we find the constraints to read $|g_3|^2 = 1$ and
\bea
\label{c1}
g_2 \bar{g}_1 + g_1 \bar{g}_2 + g_3 \bar{g}_1 + g_1 \bar{g}_3 + 
h_0 (|g_2|^2 + |g_1|^2) &=& 0,\\
\label{c2}
g_4 \bar{g}_5 + g_5 \bar{g}_4 + g_3 \bar{g}_5 + g_5 \bar{g}_3 +
h_0 (|g_4|^2 + |g_5|^2) &=& 0,\\
\label{c3}
g_3 \bar{g}_2 + g_4 \bar{g}_2 + g_5 \bar{g}_1 + g_4 g_3 +
h_0 (g_4 \bar{g}_1 + g_5 \bar{g}_2 ) &=& 0,\\
\label{c4}
g_1 \bar{g}_5 + g_2 \bar{g}_4 + g_2 \bar{g}_3 + g_3 \bar{g}_4 +
h_0 (g_1 \bar{g}_4 + g_2 \bar{g}_5 ) &=& 0.
\eea

Finally, let us briefly comment on a plausible CFT interpretation of $\langle Y_{i\bar{j}} \rangle$. In the simple case of the 2-strand braid. Taking the trace of $Y_{i\bar{j}}$, we end up with the two-by-two matrix 
\be
\label{conformBlock}
\langle Y_{i\bar{j}} \rangle= \left( \begin{array}{cc} Z \left( \bigcirc \cup \bigcirc \right) & Z \left( \bigcirc \right)  \\
 Z\left( \bigcirc \right) & Z \left( \bigcirc \cup \bigcirc \right) \end{array} \right)
\ee
Further, we can express it more suggestively as $\langle Y_{i\bar{j}} \rangle = \langle \phi_i | \phi_j \rangle$ with the inner product isomorphic to joining the punctures in region $R_\mathcal{B}$. Then upon orthonormalization of the matrix $\langle Y_{i\bar{j}} \rangle$, it was pointed out in \cite{Fradkin1} that the orthonormal states can be identified as the conformal blocks associated with the trivial and adjoint representation that appear in the fusion of the fundamental and anti-fundamental states, so \eqref{conformBlock} is essentially the fusion matrix of the CFT that lives on the boundary of the original 3-ball. It would be interesting to furnish a similar interpretation for a general $m$-strand braid and with other choices of $R_{\mathcal{B}}$.


\section{Entanglement measures and Jones polynomials}
\label{sec:Jones}

In this section, we will compute the topological entanglement entropy measures and demonstrate that entanglement measures are generally sensitive to the braiding of the Wilson lines when we trace out portions of $S^2$ containing only a proper subset of the punctures. As mentioned earlierWe perform our computations by gluing copies of punctured discs following the replica method to obtain the auxiliary link of which Jones polynomial yields the Renyi entropy.

\subsection{Single-strand braids connecting a pair of quasi-particles }

\begin{figure}[h]
\centering
\includegraphics[scale=0.6]{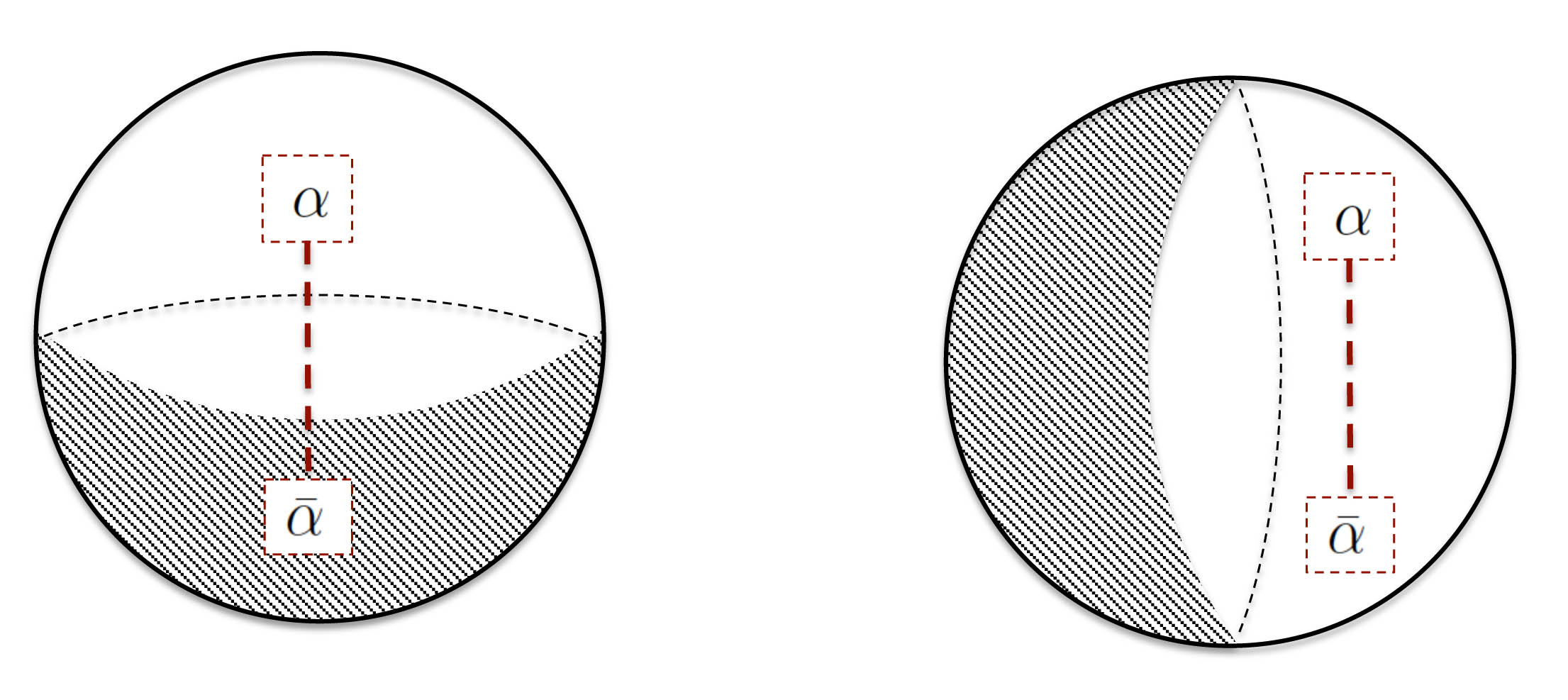}
\caption{In the left diagram, $\mathcal{R}_B$ encloses an quasi-hole while it encloses none in the right diagram. The former has the VEV of an unknotted Wilson loop as the Renyi entanglement entropy
but not the latter.}
\label{Fig7}
\end{figure}

First, let us consider a single Wilson line connecting two punctures on the sphere. Let $\mathcal{R}_B$ denote the region being traced out when we compute the entanglement entropies, as sketched in Fig. \ref{Fig7}.

We find that the entanglement entropy depends on the choice of $\mathcal{R}_B$ as follows.
\be
S_{EE} =\begin{cases}
  \text{Log} \, S_{j0}, & R_{\mathcal{B}} = \{ \bar{\alpha} \}\cr    
  \text{Log} \, S_{00}, & R_{\mathcal{B}} = \{ 0 \}\\       
\end{cases}
\ee
where $j$ denotes the representation (spin) of the Wilson line. We performed this computation by filling up the interior of $S^2$ to obtain a 3-ball and then 
gluing the appropriate discs following the surgery procedure that implements $\text{Tr} ( \rho^n )$. When $R_{\mathcal{B}} = \{ \bar{\alpha} \}$, the auxiliary link $\mathcal{L}_n$ is the unknot independent of $n$. This implies $S_{EE} = S^{(n)}_R = \text{Log}\, S_{0\frac{1}{2}}$. When $R_{\mathcal{B}} = \{0\}$, $\mathcal{L}_n$ is the split union of $n$ unknots which yields 
\be
S^{(n)}_R = \frac{1}{1-n}\text{Log} \left(  (2 \cos (\frac{\pi}{k+2}))^{n-1}/S_{0\frac{1}{2}} \right) = \text{Log} \left( \sqrt{\frac{2}{k+2}} \sin (\frac{\pi}{k+2}) \right) = \text{Log} (S_{00})
\ee
where we observe that $S_{00}$ is the Chern-Simons partition function on $S^3$ (without insertion of any Wilson loops). 

This result holds for other TQFTs, with $S_{j0}$ and $S_{00}$ to be replaced by the partition functions of the theory with and without an unknotted Wilson loop respectively. Clearly, in the case where 
$R_{\mathcal{B}} = \{\alpha, \bar{\alpha}\}\,\,\text{or}\,\, \{0\}$, the entanglement measures are insensitive to the Wilson line which intuitively does not `entangle' the degrees of freedom living on both regions.

\subsection{ 2-strand braids }

We now consider a 2-strand braid connecting two pairs of quasi-particles. 
A generic braid word is of the form $\sigma_1^m, m\in \mathbb{Z}$. For example, when $m=\pm 2$, this yields the Hopf link. It turns out that all choices of $R_\mathcal{B}$ 
give rise to Renyi and entanglement entropies which do not distinguish the braided pair with the
unbraided one, apart from the cases of $R_\mathcal{B} = \{ \alpha_1, \bar{\alpha}_1 \}$ or
$R_\mathcal{B} = \{ \alpha_2, \bar{\alpha}_2 \}$. 

\begin{figure}[h]
\centering
\includegraphics[scale=0.5]{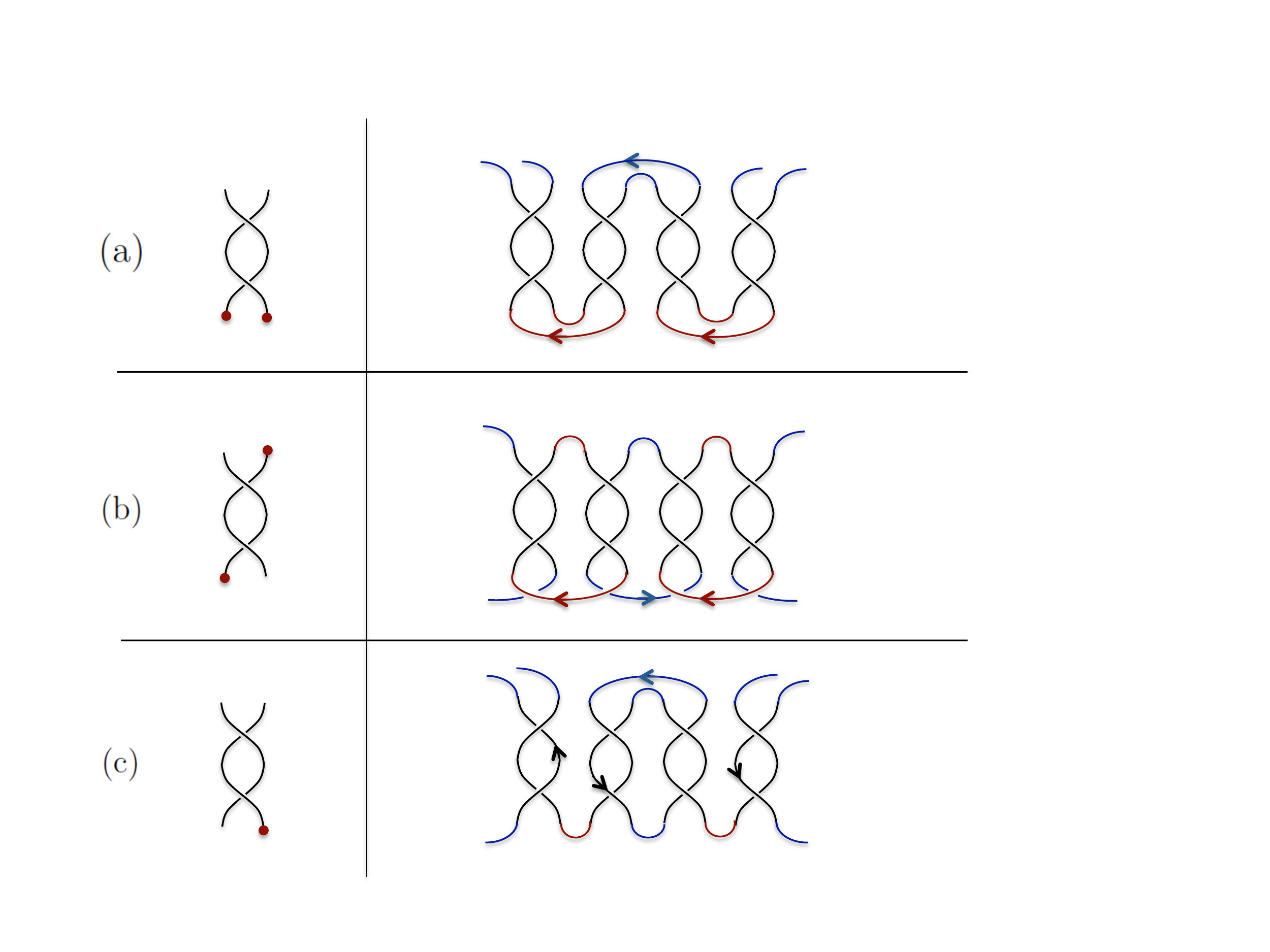}
\caption{In this figure, we sketch construction of the auxiliary link $\mathcal{L}_2$ associated with $\text{Tr}\rho^2$ for three choices of $R_{\mathcal{B}}$. The open ends on the left and right should be identified. They turn out to be identical to the unbraided case. The auxiliary link $\mathcal{L}_n$ is the split union of two unknots for (a) and (b), independent of $n$ whereas it is the split union of $n+1$ unknots in (c). }
\label{Fig8}
\end{figure}

In the following, for definiteness we will perform calculations in $SU(2)$ Chern-Simons theory with the braided Wilson lines in the fundamental. As noted earlier, the writhe number of the link $\mathcal{L}_{n}$ is always zero. We compute the Kauffman bracket polynomial in variable $A$ which is equivalent to the Jones polynomial (say in $t$) since the writhe number is zero\footnote{We recall some basic definitions slightly more formally. The Kauffman bracket is a function which sends link diagrams to Laurent polynomials with integer coeffcients, so it maps the diagram $D$ to $\langle D \rangle \in \mathbb{Z} [ A^{-1}, A ]$. It
is related to the Jones polyomial of the same link $L$ by $V_{Jones}(L) = \left( (-A)^{-3w(D)} \langle D \rangle \right)_{\sqrt{t} = A^{-2}} \in \mathbb{Z} [t^{-1/2}, t^{1/2}]$ with $D$ being some oriented link diagram and$w(D)$ its writhe number. For the auxiliary link, we saw earlier that it is fully amphichiral and has always vanishing writhe number, so the bracket of such a link is the same as its Jones polynomial. We also take all knots and links in this paper to be in the standard framing.}, with
\be
\label{skeinR}
A = i e^{-\frac{i \pi }{2(k+2)}} = t^{-1/4}. 
\ee
We made this identification by comparing \eqref{skeinR} with the Chern-Simons VEV for the unknot and the split union of 2 unknots as follows. In Chern-Simons theory, the partition functions for these two Wilson loops read 
$Z(S^3;\bigcirc) = S_{0 \frac{1}{2}}, Z(S^3;\bigcirc \cup \bigcirc) = S^2_{0 \frac{1}{2}}/S_{00}$ and thus their ratio reads $S_{0 \frac{1}{2}}/S_{00} = 2 \cos \frac{\pi}{k+2}$ which is consistent with the skein relations for the Jones or Kauffman bracket polynomial if we adopt \eqref{skeinR}.
Traces of the density matrix and its powers are equivalent to the partition function of Chern-Simons theory with the auxiliary link in $S^3$. 
In the following, we shall compute the entanglement measures for the case where $\mathcal{R}_B = \{ 
\alpha_2, \bar{\alpha}_2 \}$. This essentially reduces to computing the Jones polynomial for the auxiliary link.

\subsubsection{Two overcrossings}
We begin with the case of two overcrossings. In Fig. \ref{AuxTwo} and Fig. \ref{AuxHopf2n} below, we sketch the construction of the auxiliary link.

\begin{figure}[h]
\centering
\includegraphics[width=40mm]{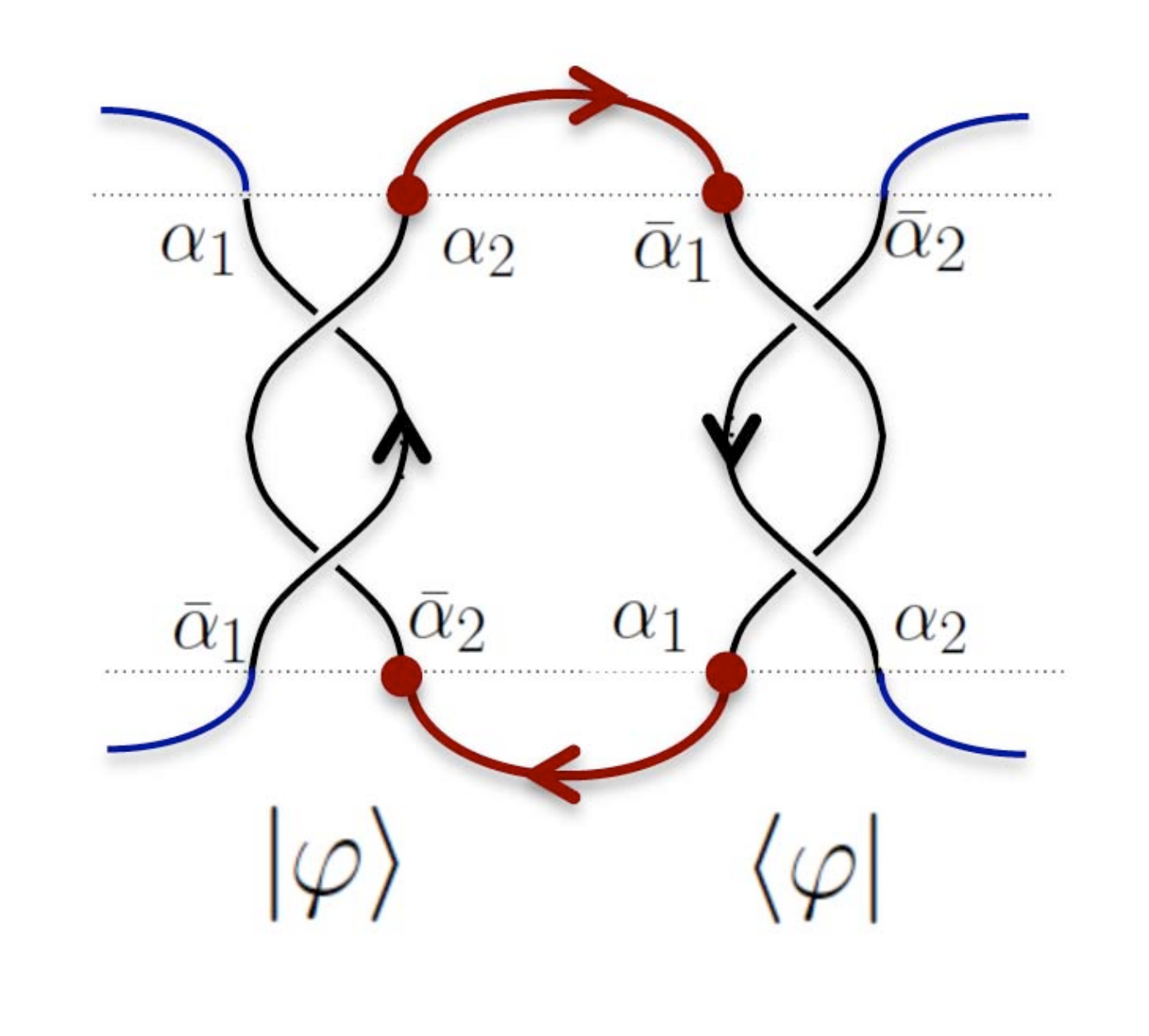}
\caption{In the above, we sketch the diagram associated with $\rho$ for the 2-strand braid with two crossings and taking
$\mathcal{R}_B = \{\alpha_2, \bar{\alpha}_2 \}$. 
If we take the trace and identify the open ends, we obtain $\mathcal{L}_1$ which is
the disjoint union of two unknots. }
\label{AuxTwo}
\end{figure}

\begin{figure}[h]
\centering
\includegraphics[width=120mm]{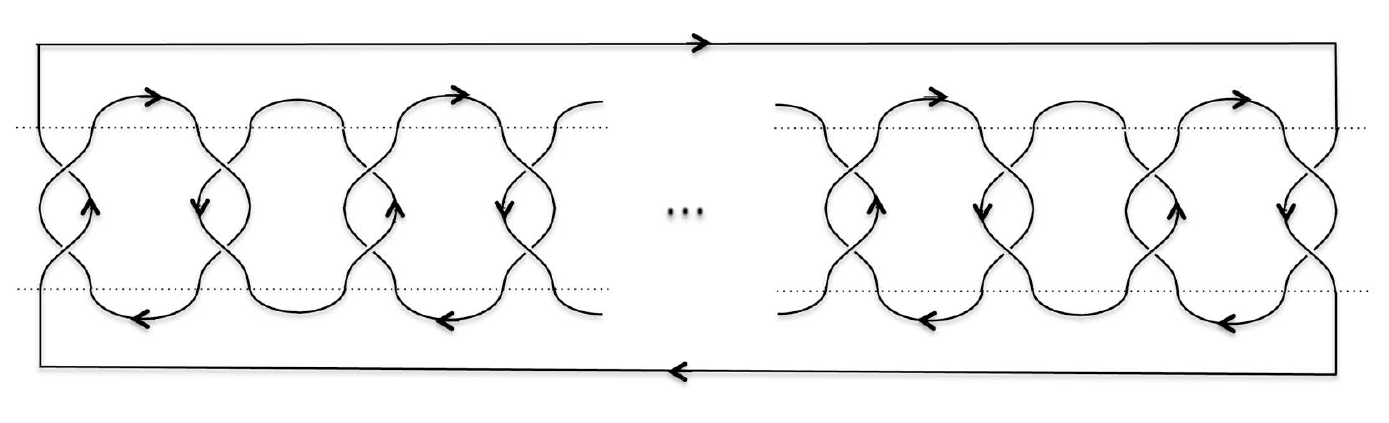}
\caption{The auxiliary link $\mathcal{L}_n$ for the 2-strand braid with two crossings which we find to be a $2n$-component link.}
\label{AuxHopf2n}
\end{figure}

Using the bracket skein relations recursively, it is straightforward to compute the polynomial which we find to be 
\be
\label{twostrandRenyi}
\langle \mathcal{L}_{n} \rangle = 
\text{Tr}\, \rho^n = -\frac{ (A^4 + A^{-4})^{2n} }{A^2 + A^{-2}} + (-1)^{n+1} (A^2 - A^{-2})^{2n}
\frac{1+A^4+A^{-4} }{A^2 + A^{-2}} 
\ee
 This turns out to be a special case of more general formula that we will derive shortly in the next section for which we reserve its detailed derivation.  We also present an independent check of this formula in Appendix A. Now from \eqref{Renyi}, \eqref{entanglemententropy} and \eqref{twostrandRenyi}, we compute the entanglement measures to be
\bea
S^n_{Renyi} &=& \text{Log}\, S_{\frac{1}{2}0} + \frac{1}{1-n}\text{Log} \left[ \frac{2^{n-1}}{\cos^{n+1} \frac{\pi}{k+2}} \left(  \cos^{2n} \frac{2\pi}{k+2} + \sin^{2n} \frac{\pi}{k+2} (1+ 2\cos \frac{2\pi}{k+2}) \right) \right] \cr
S_{EE} &=& \text{Log} S_{\frac{1}{2}0} - \frac{1}{2}\sec^2 \frac{\pi}{k+2} 
\left[   
\text{Log} \frac{2\cos^2 \frac{2\pi}{k+2}  }{\cos \frac{\pi}{k+2} } + \cos \frac{2\pi}{k+2} \left(
\text{Log} \frac{2\sin^2 \frac{\pi}{k+2} }{\cos \frac{\pi}{k+2} } \right)  
\right] \nonumber
\eea
Denoting $\delta S_{EE}$ to be the change in the entanglement entropy relative to that in the unbraided case,
in the large $k$ limit, we find
$$
\lim_{k\rightarrow \infty} \delta S_{EE} = -\text{Log} (2) + x^2 \left( \frac{7}{2} - 6\text{Log} (x) \right) + \ldots, \qquad x \equiv 
\frac{\pi}{k+2}.
$$
We will see shortly that this is a common limiting behavior of $\delta S_{EE}$ for even number of crossings.


\subsubsection{Jones polynomial of the auxiliary link for 2-strand braid}

\begin{figure}[h]
\centering
\includegraphics[scale=0.7]{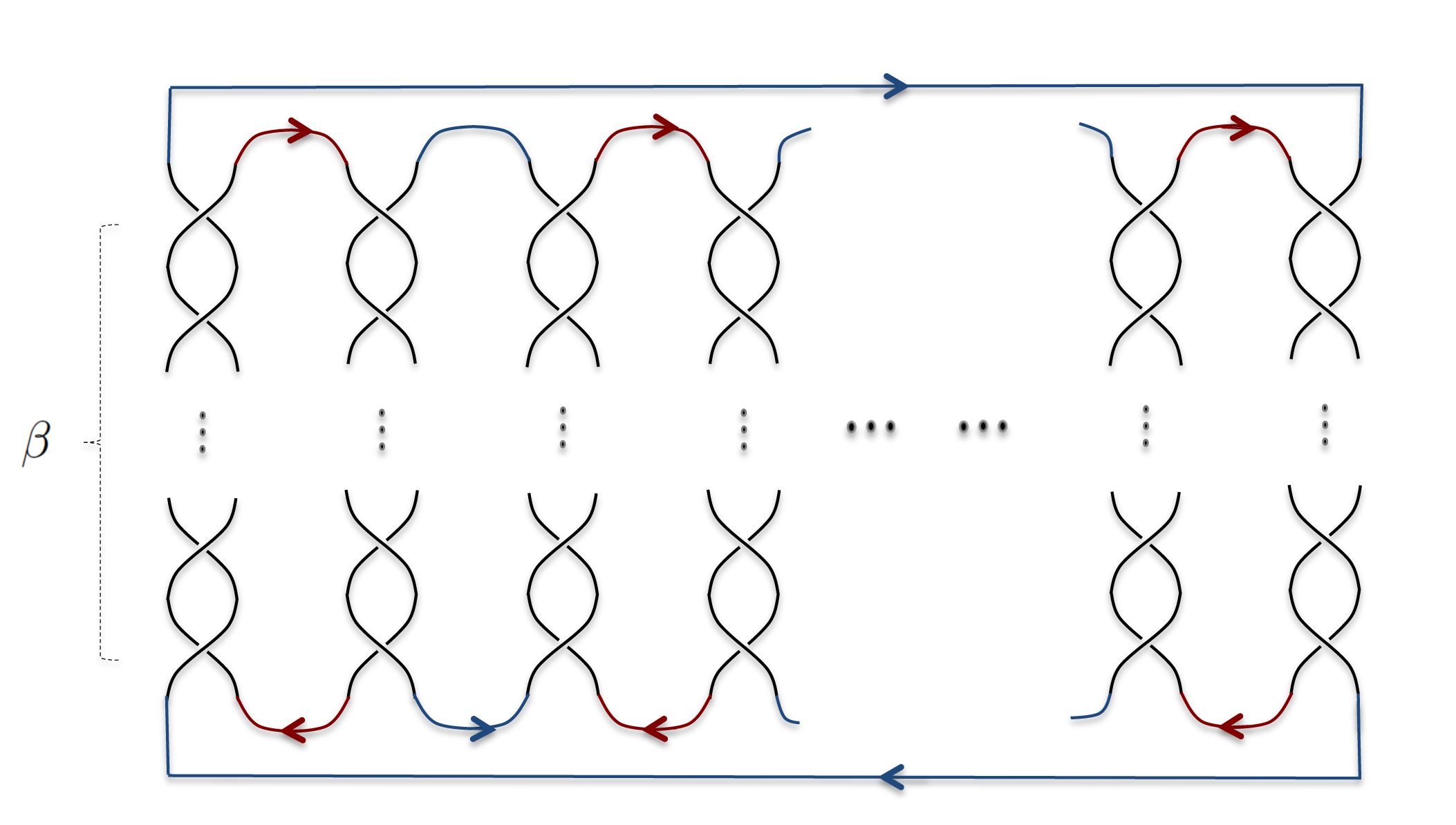}
\caption{The auxiliary link $\mathcal{L}_n$ for the 2-strand braid with $\beta$ crossings which we find to be a Pretzel link of the form indicated in \eqref{pretzel}. For $\beta$ even, it is a two-component link, whereas for an odd $\beta$, it is a $2n$-component link.}
\label{AuxBetaHopf}
\end{figure}

Below, we explore the case
of $\beta$ over-crossings, with $\beta$ being an arbitrary integer.
From the gluing procedure we find that the auxiliary link $\mathcal{L}_n$ 
is a member of a certain class of Pretzel links\footnote{See for example \cite{Landvoy} for some other general results on the Jones polynomials of Pretzel links.} as illustrated in Fig. \ref{AuxBetaHopf}.
These Pretzel links are of the form
\be
\label{pretzel}
P\left( \beta, -\beta, \beta, -\beta, \ldots,\beta,-\beta \right)
\ee
where there are $2n$ tassel with crossings alternating between $\pm \beta$ and the writhe number vanishes.

We begin by simplifying just one of the tassels which has $\beta_i$ crossings by
using \eqref{usefulresult} derived in Section \ref{universal} and resolving the crossings to obtain
\be
\label{betaRec}
P( \ldots, \beta_{i-1}, \beta_i, \beta_{i+1},  \dots ) = A^{\beta_i} P( \ldots, \beta_{i-1}, 0, \beta_{i+1},  \dots ) + 
f(\beta_i) P (\ldots, \beta_{i-1}, \beta_{i+1}, \ldots ) 
\ee
where $f(\beta) \equiv \frac{1}{h_0} \left( -A^{\beta} + (-\bar{A}^3)^\beta  \right)$. 
From \eqref{betaRec}, it is useful to note a Corollary
\be
\label{coroll}
P( \ldots, \beta, 0 ) = (A^\beta h_0 + f(\beta)) P(\ldots, 0 )
\ee
which follows from the fact that a disjoint union with an unknot gives a multiplicative factor of $h_0$ to the polynomial. From \eqref{betaRec} and \eqref{coroll}, it is then straightforward to obtain
\be
P(\beta_1, \beta_2) = \frac{1}{h_0} \left(   \prod^2_{j=1} (A^{\beta_j} h_0 + f(\beta_j) ) + (h_0^2 -1)f(\beta_j) \right)
\ee
and similarly 
\be
P(\beta_1, \beta_2, \beta_3) = \frac{1}{h_0} \left(   \prod^3_{j=1} (A^{\beta_j} h_0 + f(\beta_j) ) + (h_0^2 -1)f(\beta_j) \right)
\ee
Indeed, from \eqref{betaRec},\eqref{coroll} and by induction,
\bea
P(\beta_1, \beta_2, \ldots, \beta_m) &=& A^{\beta_m} P(\beta_1, \beta_2, \ldots, \beta_{m-1},0) + f(\beta_m) P(\beta_1, \beta_2, \ldots, \beta_{m-1}) \cr
&=&A^{\beta_m} \prod_{j=1}^{m-1} ( A^{\beta_j}h_0 +  f(\beta_j) ) + \frac{f(\beta_m)}{h_0} 
( \prod_{j=1}^{m-1} ( A^{\beta_j}h_0 +  f(\beta_j)  + (h_0^2 -1) \prod_{j=1}^{m-1} f(\beta_j) ) \cr
&=&\frac{1}{h_0} \left(   \prod^m_{j=1} (A^{\beta_j} h_0 + f(\beta_j) ) + (h_0^2 -1)f(\beta_j) \right)
\eea
For our purpose, we are interested in $m=2n$, with $\beta_j = \pm \beta$ with $+$ sign for even $j$ and $-$ sign for odd $j$. After some algebra we find that the polynomial reads
\be
\langle \mathcal{L}_{n} (\beta) \rangle = \frac{1}{h_0} \left(   
\left| A^\beta \left( A^2 + \bar{A}^2 \right) + (-1)^\beta A^{2-3\beta} \frac{1-(-A^4)^\beta}{1+A^4} \right|^{2n} + \left( (A^2+\bar{A}^2)^2-1 \right) | A^{2-3\beta} \frac{1-(-A^4)^\beta}{1+A^4} |^{2n} \right)
\ee
The entanglement measures read
\be
S^{(n)}_{R} = \text{Log}\, S_{\frac{1}{2}0} + \frac{1}{1-n}\text{Log} \left(\frac{(f)^n + g(h)^n }{h_0^{n+1}}   \right)
\ee
where $h_0 = 2 \cos \frac{\pi}{k+2}, g = 4\cos^2 \frac{\pi}{k+2} - 1 = h_0^2 - 1$.
For $\beta$ even, we have 
\be
f  = 4\cos^2 \frac{\pi}{k+2} + \frac{\sin^2 \frac{\pi \beta}{k+2}}{\cos^2 \frac{\pi}{k+2}} - 4 \sin^2 \frac{\pi \beta}{k+2},\,\,\,
h = \frac{\sin^2 \frac{\pi \beta}{k+2}}{\cos^2 \frac{\pi}{k+2}}.
\ee
whereas for $\beta$ odd, we have
\be
f  = 4\cos^2 \frac{\pi}{k+2} + \frac{\sin^2 \frac{\pi \beta}{k+2}}{\cos^2 \frac{\pi}{k+2}} - 4 \cos^2 \frac{\pi \beta}{k+2},\,\,\,
h = \frac{\cos^2 \frac{\pi \beta}{k+2}}{\cos^2 \frac{\pi}{k+2}}.
\ee
The entanglement entropy reads 
\be
S_{EE} = \text{Log}\, S_{\frac{1}{2}0} - \frac{1}{h_0^2} \left(  f \,\,\text{Log} \left(\frac{f}{h_0} \right) + g
h \,\,\text{Log} \left(\frac{h}{h_0} \right) \right)
\ee
In Fig. \ref{see1} and \ref{see2}, we display the dependence of $S_{EE}$ on the Chern-Simons coupling $k$. 
\\[5ex]
\begin{figure}[h]
\centering
\includegraphics[width=100mm]{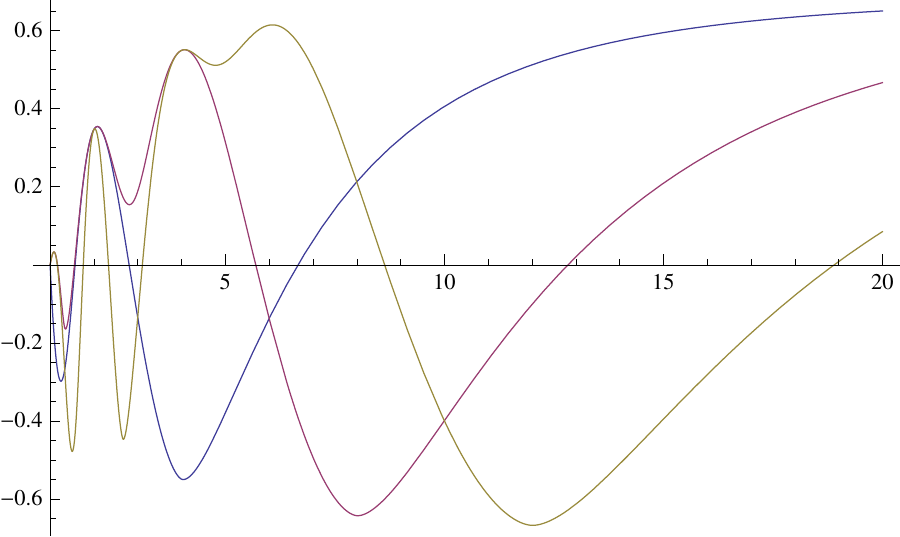}
\put(2,82){$k$}
\put(-280,180){$\delta S_{EE}$}
\put(-40,62){$\beta=7$}
\put(-40,119){$\beta=5$}
\put(-40,172){$\beta=3$}
\caption{Plot of entanglement entropy for 3,5,7 overcrossings. The odd overcrossings's entropies all tend to Log(2) at infinite $k$. }
\label{see1}
\end{figure}
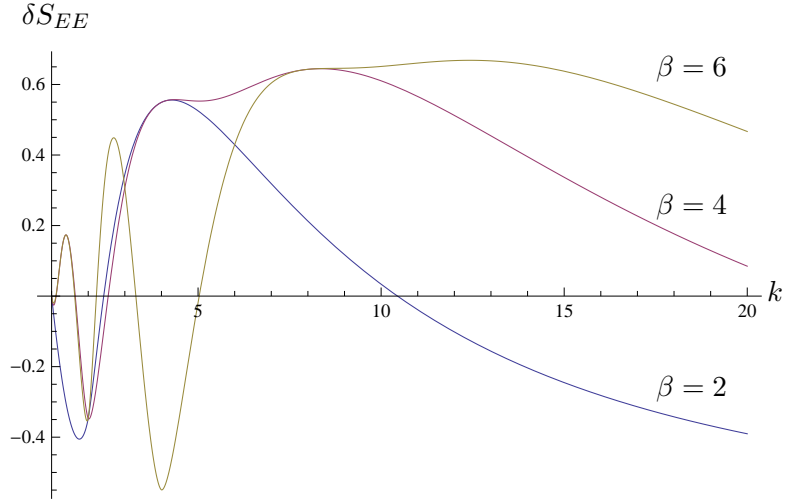
\begin{figure}[h]
\centering
\includegraphics[width=100mm]{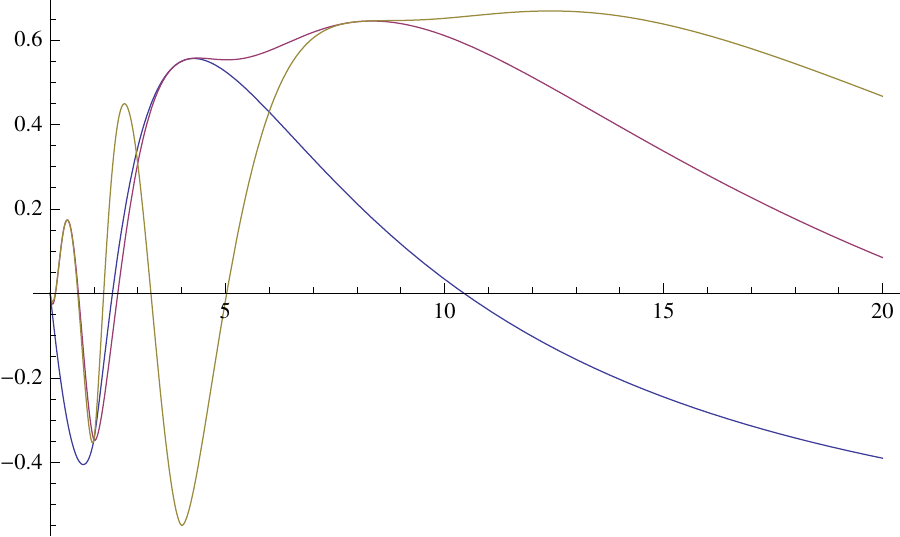}
\put(2,75){$k$}
\put(-280,180){$\delta S_{EE}$}
\put(-40,39){$\beta=2$}
\put(-40,108){$\beta=4$}
\put(-40,160){$\beta=6$}
\caption{Plot of entanglement entropy for 2,4,6 overcrossings. The odd overcrossings's entropies all tend to -Log(2) at infinite $k$. }
\label{see2}
\end{figure}

Let us end this section with some concluding remarks. 
We saw that a generic choice of $R_{\mathcal{B}}$ 
does not always lead to entanglement measures which distinguish the braided from the unbraided case. Taking $R_{\mathcal{B}} = \{ \alpha_2, \bar{\alpha}_2 \}$, we find that the auxiliary link is a family of  links parametrized by $\beta, n$. We computed its Jones polynomial and thus the 
entanglement measures of which deviations from those of the trivial case are 
manifest in $\delta S_{EE}$ being a function in $k$ and $\beta$. At large $k$,
$\delta S_{EE} \rightarrow \mp \text{Log} 2$ which is a simple indicator of the parity of the number of crossings.


\subsection{3-strand braid: the connected sum of two Hopf links}

\begin{figure}[h]
\centering
\includegraphics[scale=0.25]{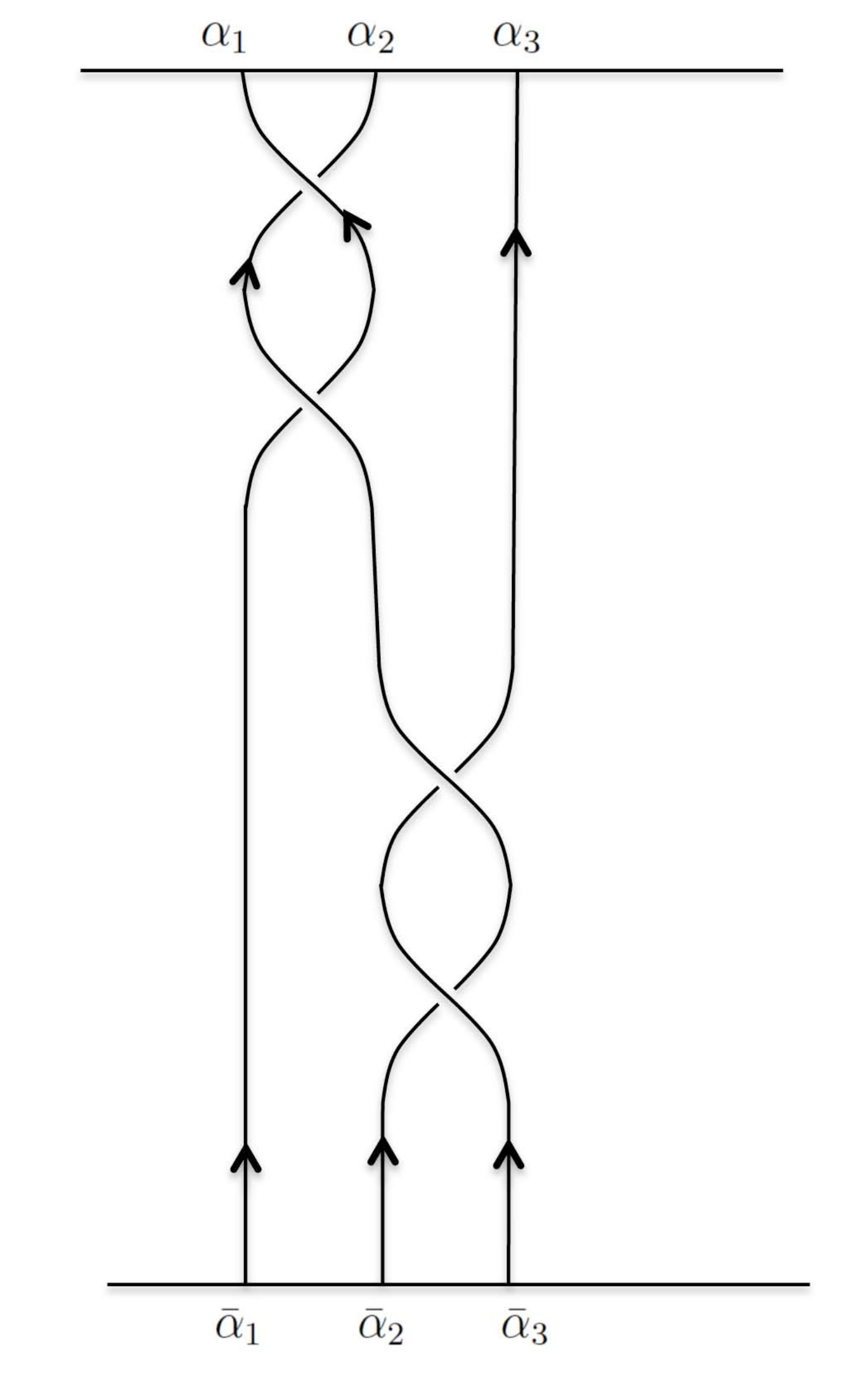}
\caption{Braid presentation for the connected sum of two hopf links. }
\label{FigBraidComposite}
\end{figure}

In the following we consider the case of the connected sum of two Hopf links which has the following 3-strand braid presentation as shown in Figure \ref{FigBraidComposite}.
We find that if we choose the region $R_B = \{ \bar{\alpha}_3, \alpha_2, \alpha_3 \}$ then the 
Renyi entropy distinguishes it with the disjoint union of an unknot and a Hopf link or that of three unknots. 
We begin with the auxiliary link of which
diagram is shown below in Fig. \ref{AuxHump}. This turns out to be a $(2M+1)$-component link, as depicted in Fig. \ref{AuxHump}.\footnote{In this section, we use the index $M$ to denote the power index of the density matrix and its auxiliary link $\mathcal{L}_M$ for notational convenience as we choose to use the symbol $n$ as a dummy index.} 
\begin{figure}[h]
\centering
\includegraphics[scale=0.8]{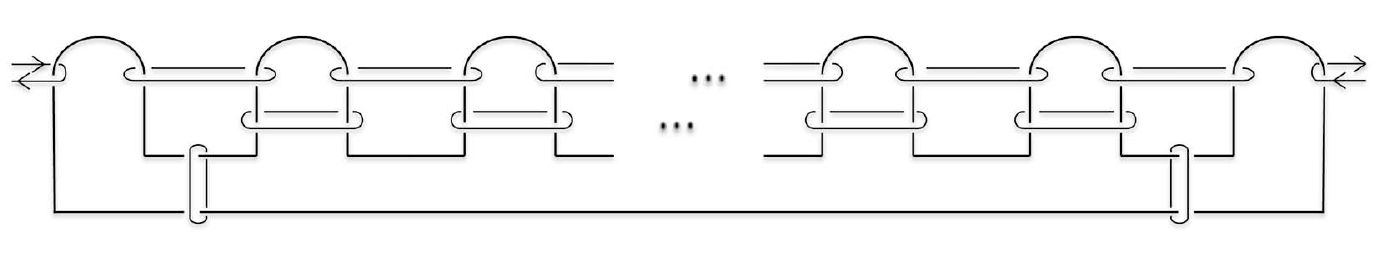}
\caption{The auxiliary link for the connected sum of two hopf links has $2M+1$ components. The open lines at each end are identified correspondingly.}
\label{AuxHump}
\end{figure}
\newpage
We find the following skein relations in Fig.\ref{basicskeins} useful for resolving the crossings on the `humps' (see Fig. \ref{AuxHump}). This reduces the link diagram eventually to the one shown on the LHS of Fig. \ref{CompSkein3}.
\begin{figure}[h]
\centering
\includegraphics[scale=0.6]{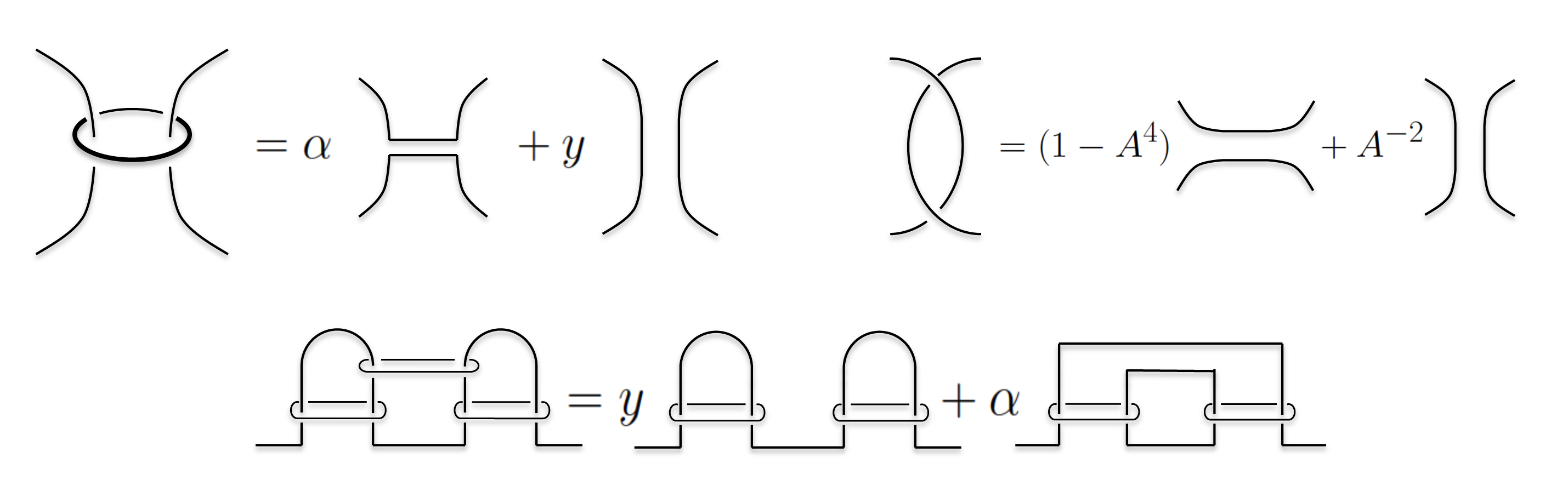}
\caption{We note some recursive skein relations which we find to be useful for computing the Jones polynomial of the auxiliary link. These are invoked to reduce the auxiliary link in Fig. \ref{AuxHump} to the link on the LHS of Fig. \ref{CompSkein3} below.}
\label{basicskeins}
\end{figure}
\begin{figure}[h]
\centering
\includegraphics[scale=0.8]{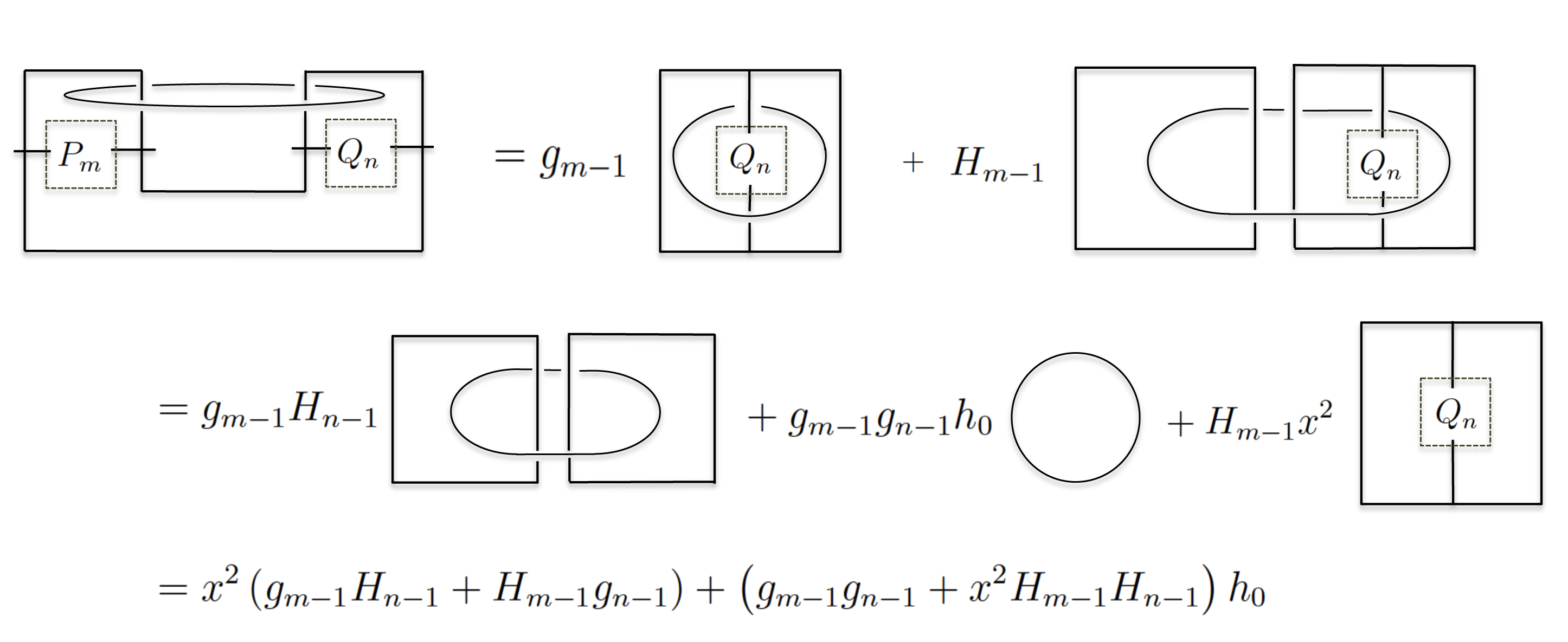}
\label{fig:a}
\end{figure}
\begin{figure}[H]
\centering
\includegraphics[scale=0.6]{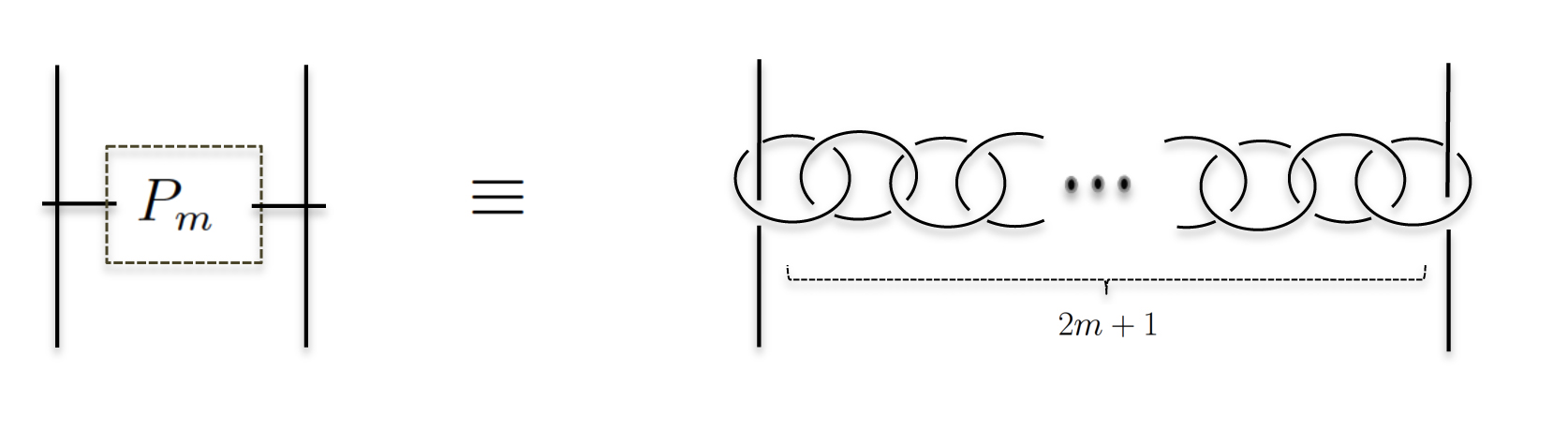}
\put(-350,85){where we have denoted}
\caption{This figure depicts how we use a simple set of skein relations to compute the Jones polynomial of the auxiliary link in Fig. \ref{AuxHump}.} 
\label{CompSkein3}
\end{figure}
\newpage
\newpage

\begin{figure}[h]
\centering
\includegraphics[scale=0.9]{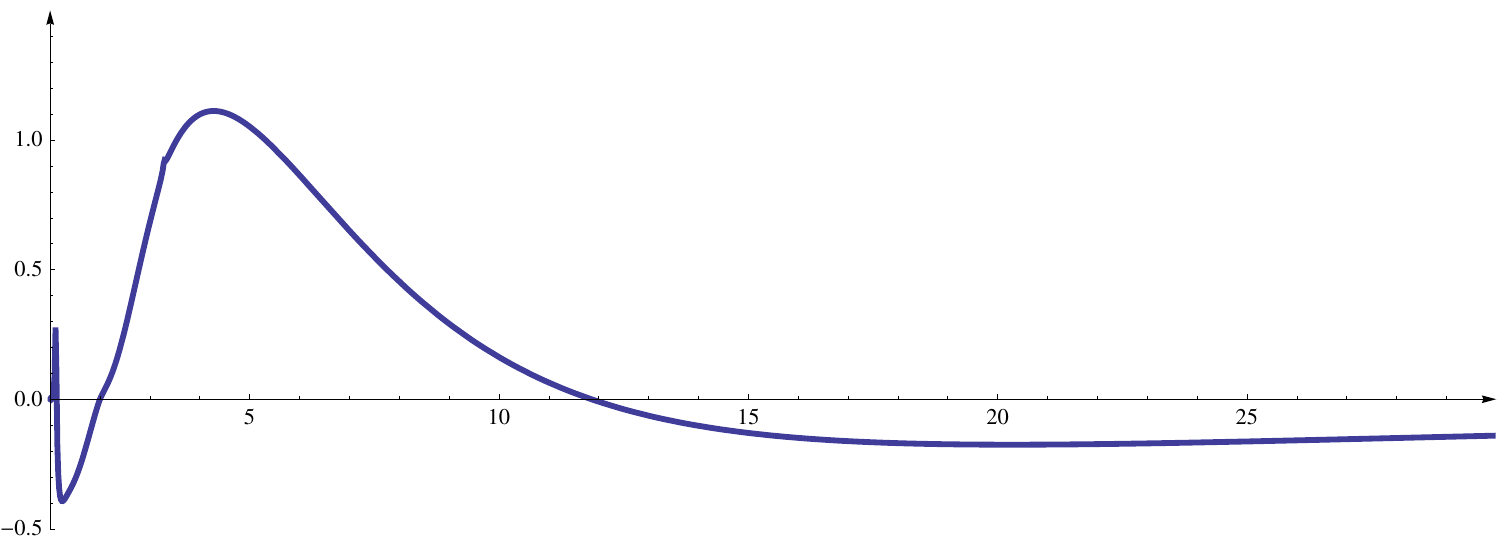}
\put(-2,32){$k$}
\put(-400,142){$\delta S_{EE}$}
\caption{Graph showing how $\delta S_{EE}$ varies as a function of Chern-Simons level $k$. It tends to zero as $k$ approaches infinity. }
\label{see3}
\end{figure}

We note that in Fig. \ref{CompSkein3}, the various parameters are defined as
\bea
\label{param}
&&x = -A^4 - \bar{A}^4, \,\,\, \alpha = |1-A^4|^2, \,\,\, y = -A^6 - \bar{A}^6,\,\,\, h_0 = - A^2 - \bar{A}^2,\,\,\, r = \alpha/x^2, \cr
&& H_k= |1-A^4|^{2k+2},\,\,\, g_k = yx^{2k} \left(  \frac{r^{k+1} -1}{r-1} \right),\,\,\, \mathcal{P}_k =
h_0 H_k + g_k,\cr
&& \mathcal{B}_{kl} = x^2 \left( g_k H_l + g_l H_k \right) + h_0 \left( g_k g_l + x^2 H_k H_l \right).
\eea
After some algebra, we obtain the Chern-Simons VEV to read 
\bea
\langle  \mathcal{L}_M \rangle &=& 
\sum_{l=1}^{M-4} \sum_{r=1}^{M-3-l} \mathcal{B}_{lr} \sum^{M-3-l-r}_{k=0} \left(  \begin{array}{c} M-3-l-r \\ k \end{array} \right) (yh_0)^{M-3-l-r-k} (\alpha^k \mathcal{P}_k ) y^2 \alpha^{l+r} \cr
&+& y \alpha^{M-2} \sum^{M-2}_{l=0} \mathcal{B}_{l,M-2-l} + 2 \sum^{M-3}_{l=1} \alpha^l \mathcal{B}_{l0} \sum^{M-3-l}_{k=0} \left( \begin{array}{c} M-3-l \\ k \end{array} \right) (yh_0)^{M-3-l-k} \alpha^k \mathcal{P}_k y^2 \cr
\label{linkHopfP1}
&+& y^2 \mathcal{B}_{00} \sum^{M-3}_{k=0} \left(  \begin{array}{c} M-3 \\ k \end{array} \right) (yh_0)^{M-3-k} (\alpha^k \mathcal{P}_k  ) + h_0 \mathcal{P}_{M-1} \alpha^{M-1}
\eea
where $\alpha = |1-A^4|^2$. As a consistency check, for every term that appears in \eqref{linkHopfP1}, the powers of $y$ and $\alpha$ should sum up to $M-1$. We note that 
$\mathcal{P}_k \neq \mu^k$ for some $\mu = \mu (A)$, but it can be expressed as 
\bea
\mathcal{P}_m &=& h_0 H_m + g_m \cr
&=& h_0 |1- A^4 |^2 ( |1-A^4|^2 )^m + \frac{yr}{r-1} (x^2 r)^m + \frac{y}{1-r}(x^2)^m \cr
\label{Psum}
&\equiv& \sum^2_{i=1} C_i p^m_{i}
\eea
where
$$
C_1 \equiv h_0 \alpha + \frac{yr}{r-1},\,\, C_2 \equiv \frac{y}{1-r},\,\, p_1 \equiv \alpha,\,\, p_2 \equiv x^2.
$$
Substituting \eqref{Psum} into \eqref{linkHopfP1}, we can perform the sum over the dummy index $k$ in each term to obtain 
\bea
\langle \mathcal{L}_n \rangle &=& \sum^2_{i=1} C_i \sum^{M-4}_l \sum^{M-3-l}_{r=1} \mathcal{B}_{lr} ( yh_0 + \alpha p_i )^{M-3-l-r} y^2 \alpha^{l+r} + h_0 \mathcal{P}_{M-1} \alpha^{M-1} \cr
&+& \sum^2_{i=1} C_i y^2 \mathcal{B}_{00} (yh_0 + \alpha p_i )^{M-3} + y \alpha^{M-2} \sum^{M-2}_{l=0} \mathcal{B}_{l,M-2-l} \cr
\label{linkHopfP2}
&+& 2 \sum^2_{i=1} C_i \sum^{M-3}_{l=1} \alpha^l \mathcal{B}_{l0} (yh_0 + \alpha p_i )^{M-3-l} y^2.
\eea
We note that after some simplification, $\mathcal{B}_{kl}$ can be written in the form 
\be
\label{Bdefin}
\mathcal{B}_{kl} = \sum^2_{m,n=1} b_{mn} p^k_m p^l_n
\ee
where 
$$
b_{11} = 2 \alpha^2 C_2 + h_0 (C^2_2 r^2 + x^2 \alpha^2 ), b_{12} = b_{21} = -h_0 r C^2_2 - x^2 \alpha C_2, b_{22} = h_0 C^2_2. 
$$
Substituting \eqref{Bdefin} into \eqref{linkHopfP2}, we obtain 
\bea
\langle \mathcal{L}_M \rangle &=& y^2 
\sum^2_{i,k,l=1} C_i \sum^{M-4}_{l=1} \sum^{M-3-l}_{r=1} b_{kl} p^l_m p^r_n (yh_0 + \alpha p_i )^{M-3-l-r}\alpha^{l+r} + h_0 \mathcal{P}_{M-1} \alpha^{M-1} \cr
&+& \sum^2_{i=1} C_i y^2 \mathcal{B}_{00} (yh_0 + \alpha p_i )^{M-3} + y \alpha^{M-2} \sum^{M-2}_{l=0} \sum^2_{m,n=1} b_{mn}p^l_m p^{M-2-l}_n \cr
&+& 2 \sum^2_{i=1} C_i  \sum^{M-3}_{l=1} \sum^2_{m,n=1} \alpha^l b_{mn} p^l_m (yh_0 + \alpha p_i )^{M-3-l} y^2 \cr
&\equiv& y^2 \sum^2_{i,m,n=1} C_i b_{mn} \sum^{M-4}_{l=1} \sum^{M-3-l}_{r=1} p^l_m p^r_n h^{M-3-l-r}_i \alpha^{l+r} \cr
&+& y^2 \mathcal{B}_{00} \sum^2_{i=1} C_i h^{M-3}_i + y \alpha^{M-2} \sum^2_{m,n=1} \sum^{M-2}_{l=0} p^l_m p^{M-2-l}_n \cr
\label{linkHopfP3}
&+& 2y^2 \sum^2_{i,m,n=1} b_{mn} C_i \sum^{M-3}_{l=1} \alpha^l p^l_m h^{M-3-l}_i + h_0 \mathcal{P}_{M-1} \alpha^{M-1}.
\eea
where we define $h_i = yh_0 + \alpha p_i$.
In this form, it is evident that we can express the Jones polynomial explicitly as a function of $M$. 
We now perform the sum over the indices $l$ and $r$ in each term in \eqref{linkHopfP3}. The various geometric sums read
\be
\label{gsum1}
\sum^{M-3}_{l=1} \alpha^l p^l_m \frac{h^{M-3}_i}{h^l_i} = h^{M-3}_i \sum^{M-3}_{l=1} \left[ \frac{\alpha p_m}{h_i} \right]^l
\equiv h^{M-3}_i \sum^{M-3}_{l=1} U^l_{mi} = h^{M-3}_i U_{mi} \frac{( U^{M-3}_{mi} -1 )}{U_{mi} -1},
\ee
\be
\label{gsum2}
\sum^{M-2}_{l=0} \left[  \frac{p_m}{p_n} \right]^l p^{M-2}_n \equiv p^{M-2}_n \sum^{M-2}_{l=0} V^l_{mn} = p^{M-2}_n \frac{V^{M-1}_{mn} -1}{V_{mn}-1}
\ee
\bea
\sum^{M-4}_{l=1} p^l_m h^{M-3-l}_i \alpha^l \sum^{M-3-l}_{r=1} \left[ \frac{p_n \alpha}{h_i} \right]^r &=&
h^{M-3}_i \sum^{M-4}_{l=1} \left[   \frac{p_m \alpha}{h_i} \right]^l \sum^{M-3-l}_{r=1} U^r_{ni} \cr
&=& h^{M-3}_i \sum^{M-4}_{l=1}  \left[   \frac{p_m \alpha}{h_i} \right]^l \frac{U_{ni}( U^{M-3-l}_{ni} -1   )  }{U_{ni} -1 } \cr
&\equiv & h^{M-3}_i \left[  \frac{U^{M-2}_{ni}   }{U_{ni} - 1}  \frac{V_{mn} ( V^{M-4}_{mn} - 1)     }{(V_{mn} - 1)}   -    \frac{U_{ni}   }{U_{ni} - 1}  \frac{U_{mi} ( U^{M-4}_{mi} - 1)     }{(U_{mi} - 1)}          \right] \nonumber \\
\eea
where
$$
U_{mi} \equiv \frac{\alpha p_m}{h_i},\,\, V_{mn} = \frac{p_m}{p_n}.
$$
Finally, assembling all the terms together, the Jones polynomial of the auxiliary link reads
\bea
\langle \mathcal{L}_M \rangle &=& y^2 \sum^2_{i,m,n=1} C_i b_{mn} h^{M-3}_i \left[  \frac{U^{M-2}_{ni}   }{ U_{ni} - 1}  \frac{V_{mn} ( V^{M-4}_{mn} - 1)}{(V_{mn} - 1)}   
- \frac{U_{ni}   }{U_{ni} - 1}  \frac{U_{mi} ( U^{M-4}_{mi} - 1)     }{(U_{mi} - 1)}   \right] \cr
&+& y^2 \mathcal{B}_{00} \sum^2_{i=1} C_i h^{M-3}_i + y\alpha^{M-2} \sum^2_{m,n=1} b_{mn} p^{M-2}_n \frac{V^{M-1}_{mn} -1}{V_{mn}-1} \cr
\label{JonesCom}
&+& 2y^2 \sum^2_{i,m,n=1} b_{mn} C_i h^{M-3}_i U_{mi} \frac{( U^{M-3}_{mi} -1 )}{U_{mi} -1} + h_0 \sum^2_{i=1} C_i (\alpha p_i)^{M-1}.
\eea
This expression is manifestly valid for $M> 4$ which we use to derive the Jones polynomial. However,
as a function in $M$, we find that it naturally extends to all lower values. To verify this, we list 
the polynomial $\langle \mathcal{L}_M \rangle$ for each lower value of $M$ below.
\bea
\label{lower4}
\langle \mathcal{L}_4 \rangle &=& h_0 \mathcal{P}_3 \alpha^3 + y^2 \mathcal{B}_{00} 
\sum_{k=0}^1 (yh_0)^{1-k}\alpha^k \mathcal{P}_k + y\alpha^2 \sum_{l=0}^2 \mathcal{B}_{l,2-l} +
2y^2 \alpha \mathcal{B}_{10}\mathcal{P}_0 y^2, \\
\label{lower3}
\langle \mathcal{L}_3 \rangle &=& 2y\alpha \mathcal{B}_{10} + h_0 \mathcal{P}_2 \alpha^2 + y^2 \mathcal{P}_0 \mathcal{B}_{00}, \\
\label{lower2}
\langle \mathcal{L}_2 \rangle &=& y\mathcal{B}_{00} +h_0 \mathcal{P}_1 \alpha,\,\,\,
\langle \mathcal{L}_1 \rangle = h^2_0
\eea
For $M=1$, only the term $h_0 \sum_i C_i (\alpha p_i)^{M-1} = h^2_0$ survives and this yields the Jones polynomial of the split union
of three unknots which is expected of $\text{Tr} \rho$. We check that we can indeed obtain \eqref{lower4}-\eqref{lower2} from \eqref{JonesCom} which can be expressed in a form manifestly appropriate for all $M \geq 1$ as follows.

\bea
\langle \mathcal{L}_M \rangle &=& 
\sum^2_{i=1} \sum_{l=0}^{M-1} \sum_{r=0}^{M-1-l} C_i h^{M-3-l-r}_i \mathcal{B}_{lr} y^2 \alpha^{l+r} 
-\sum^2_{i=1} \sum_{l=0}^{M-1} C_i h^{-2}_i \mathcal{B}_{l,M-1-l} y^2 \alpha^{M-1} \cr
&&+ \alpha^{M-2} \left(  y - y^2\sum^2_{i=1} \frac{C_i}{h_i} \right) \sum_{l=0}^{M-2} \mathcal{B}_{l,M-2-l} 
+h_0 \mathcal{P}_{M-1} \alpha^{M-1} \cr  \cr
&=& \sum^2_{i,m,n=1} \frac{C_i h^{M-3}_i y^2}{U_{ni} -1} b_{mn} \left(    U^M_{ni} \frac{V^M_{mn} - 1}{V_{mn}-1} 
- \frac{U^M_{mi}-1}{U_{mi}-1} \right) 
-\sum^2_{i,m,n=1} b_{mn} C_i h^{-2}_i y^2 (\alpha p_n)^{M-1} \frac{V^M_{mn}-1}{V_{mn}-1} \cr
\label{JfinalHopf}
&&+ \alpha^{M-2} \left(  y - y^2\sum^2_{i=1} \frac{C_i}{h_i} \right) \sum^2_{m,n=1} b_{mn}
\frac{V^{M-1}_{mn}-1}{V_{mn}-1} p^{M-2}_n
+h_0 \mathcal{P}_{M-1} \alpha^{M-1}.
\eea
From \eqref{JfinalHopf}, we can read off the Renyi entropy straightforwardly while the entanglement entropy can be simplified to read
\be
S_{EE} = -\frac{1}{h^2_0} \lim_{M\rightarrow 1} \partial_M \langle \mathcal{L}_M \rangle + \text{Log}\, 
\left( h^2_0 S_{0 \frac{1}{2}} \right),
\ee
where 
\bea
\lim_{M\rightarrow 1} \partial_M \langle \mathcal{L}_M \rangle &=& \sum^2_{i=1}
\frac{C_i y^2}{h^2_i} \sum^2_{m,n=1} b_{mn} \Bigg[  \frac{U_{ni} \text{Log} (h_i U_{ni} )}{U_{ni}-1} + 
\frac{U_{ni} (V_{mn} \text{Log} (V_{mn}) )}{(U_{ni}-1)(V_{mn}-1)}  \cr
&&-\text{Log} (\alpha p_n ) - \frac{V_{mn} \text{Log} (V_{mn})}{V_{mn}-1} 
+ \frac{\text{Log}(h_i)-U_{mi} \text{Log} (h_i U_{mi})   }{(U_{mi}-1)(U_{ni}-1)}
\Bigg] \cr
&&+ \frac{y}{\alpha}\left(  1 - y\sum^2_{i=1} \frac{C_i}{h_i}    \right) \sum^2_{m,n=1} \frac{ b_{mn} \text{Log} (V_{mn} )  }{p_n (V_{mn}-1)} +
h_0 \sum^2_{i=1} C_i \text{Log} (\alpha p_i)
\eea
For the same $R_{\mathcal{B}}$, in the case of the unbraided 3-strand, we find the auxiliary link to be split union of $2M+1$ unknots and thus
\be
\langle \mathcal{L}_M \rangle = h^{2M}_0,\qquad S_{EE} = S^{(M)}_R = \text{Log}\, S_{0 \frac{1}{2}}.
\ee 
We find that identical results hold for the case of
the split union of a Hopf link and an unknot as represented on the 3-strand braid. Thus, the difference is simply
\be
\delta S_{EE} =  -\frac{1}{h^2_0} \lim_{M\rightarrow 1} \partial_M \langle \mathcal{L}_M \rangle + \text{Log}\, 
\left( h^2_0 \right).
\ee
As a function in the Chern-Simons level $k$, and for $k$ being a positive integer, we find that 
$\delta S_{EE} = 0$ only at $k=1,2$ and also at infinity where 
for all the three cases,
\be
\lim_{k\rightarrow \infty}  \langle \mathcal{L}_M \rangle = 4^M.
\ee
Another distinguished value turns out to be $k=4$, where 
$\langle \mathcal{L}_M \rangle = 3$ for all values of $M$ and thus $\delta S_{EE} = \delta S^{(M)}_R = \text{Log}\, 3$. Fig. \ref{see3} plots a sketch of $\delta S_{EE}$ as a function in $k$. 
\\[5ex]

Let's also briefly comment on another choice
$R_{\mathcal{B}} = \{ \alpha_2, \alpha_3, \bar{\alpha}_2, \bar{\alpha}_3 \}$ where we find that the auxiliary link is the split union of the unknot and that of the 2-strand representation of the Hopf link. It thus yields the same entropy measures as the 2-strand case considered previously. Although this choice doesn't distinguish between the composite and single Hopf link (or its split union with another unknot), it separates them from the auxiliary link of the trivial braid which we find to be the split union of $3M$ unknots. Thus, from the $\delta S_{EE}$'s corresponding to these two choices of $R_{\mathcal{B}}$, we can distinguish between a Hopf link, a connected sum of two Hopf links and unknots for almost all values of $k$.


\section{Topological properties of auxiliary link from braid data}
\label{sec:Topology}


\subsection{Auxiliary link group for the 2-strand braid}

In the following we compute the link group of the auxiliary link via its Wirtinger presentation, 
taking $\mathcal{R}_B = \{ \alpha_2, \bar{\alpha}_2 \}$. Following conventional notations (see for example \cite{Lick}), we let a group generator $g_i$ associated with some $i$th segment of the link diagram represent the loop that, beginning from a base point above the planar diagram, goes straight to the $i$th over/under-passing arc, encircles it counterclockwise and returns to the base point. For each crossing we take a relator $r$ as follows. Denoting the overpass arc by $g_k$ and the underpass arc by $g_i$ as it approaches the crossing and $g_j$ as it leaves, we have $r = g_k g_i g^{-1}_k g^{-1}_j$ for an undercrossing and $r = 
g^{-1}_k g_i g_k g^{-1}_j$ for an overcrossing. When equated to the identity, the relation then asserts that the two generators corresponding to the underpassing arc are conjugate by means of the overpassing generator or its inverse depending on the sign of the crossing.

\begin{figure}[h]
\centering
\includegraphics[scale=0.2]{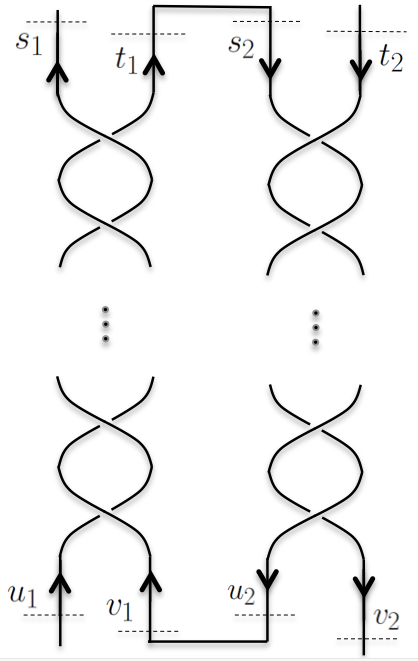}
\caption{We sketch the generators for two tassels which correspond to a factor of $\rho$. The complete auxiliary link of the 2-strand braid is sketched in Figure \ref{AuxBetaHopf}.}
\label{fund1}
\end{figure}

The auxiliary link is a Pretzel link of a certain form. For each tassel, let $s,t$ denote the generators corresponding to meridians in the link exterior that pass below the topmost two strands and $u,v$ denote
the generators passing below the bottom two strands. For convenience, we orientate the generators such that they are defined with respect to all strands being directed upwards (see Figure \ref{fund1}). 

From its Wirtinger presentation, we find the following relations prior to joining the tassels. For $\beta$
overcrossings in the $i$th tassel,
\bea
\text{For even}\,\,\beta = 2k, 
&&u_i = (t_i s_i )^{k-1}(t_i s_i t_i^{-1}) (t_i s_i)^{1-k},\,\,
v_i = (t_i s_i)^k t_i (t_i s_i)^{-k}. \cr
\label{w1}
\text{For odd}\,\,\beta = 2k+1,
&&u_i = (t_i s_i)^k t_i (t_i s_i)^{-k},\,\,
v_i = (t_i s_i)^k (t_i s_i t_i^{-1})(t_i s_i)^{-k}
\eea 
whereas for $\beta$ undercrossings, we have
\bea
\text{For even}\,\,\beta = 2k, 
&&u_i = (t_i s_i )^{-k} s_i  (t_i s_i)^{k},\,\,
v_i = (t_i s_i)^{1-k} s^{-1}_i t_i s_i (t_i s_i)^{k-1}. \cr
\label{w2}
\text{For odd}\,\,\beta = 2k+1,
&&u_i = (t_i s_i)^{-k} (s^{-1}_i t_i s_i) (t_i s_i)^k,\,\,
v_i = (t_i s_i)^{-k}  s_i (t_i s_i)^k
\eea 
Joining the $n$ tassels in the auxiliary link $\mathcal{L}_n$ implies the following $2n $ relations 
\be
\label{relations}
v_i u_{i+1} = 1, \, \forall i=1,2,\ldots, 2n,
\ee
and we identify
\be
u_{2n+1} = u_1, \, t_i = s^{-1}_{i+1}, \, s_{2n+1} = s_1.
\ee
Any one of the relations in \eqref{relations} is implied by the rest since 
we have 
\be
\prod_{j=1}^{2n} v_j u_{j+1} = 1.
\ee
We can now present the fundamental group in terms of the $2n$ generators.
\be
\pi_1 \left( S^3 - \mathcal{L}_n \right) = \langle s_1, s_2, \ldots, s_{2n} | r_1, r_2, \ldots r_{2n-1} \rangle
\ee
where each relation $r_i$ reads
\bea
\label{relationtwoeven}
\text{For even}\,\,\beta: &&
(t_i s_i)^{1-k} s^{-1}_i t_i s_i (t_i s_i)^{k-1} 
(t_{i+1} s_{i+1})^{k-1} 
t_{i+1} s_{i+1} t^{-1}_{i+1} (t_{i+1} s_{i+1})^{1-k}  = 1 \\
\label{relationtwoodd}
\text{For odd}\,\,\beta: &&
(t_i s_i)^{-k}  s_i (t_i s_i)^k  (t_{i+1} s_{i+1})^k 
t_{i+1}  (t_{i+1} s_{i+1})^{-k} = 1,
\eea

As a consistency check, let's take the simplest case of $n=1$ in which case the auxiliary link is the split union of two unknots. We have the generators $s_1, s_2$
with $t_1=s^{-1}_2, t_2 = s^{-1}_1$. We have only one relation which can be shown to be
identically satisfied for both
\eqref{relationtwoeven} and \eqref{relationtwoodd}. Since there is no quotient action
on the free group of the two generators, the fundamental group is 
$\mathbb{Z} \oplus \mathbb{Z}$ which is indeed that for the split union of two unknots.

\subsection{Auxiliary link group for the connected sum of two twist links}
\label{sec:group}
\begin{figure}[h]
\centering
\includegraphics[scale=0.5]{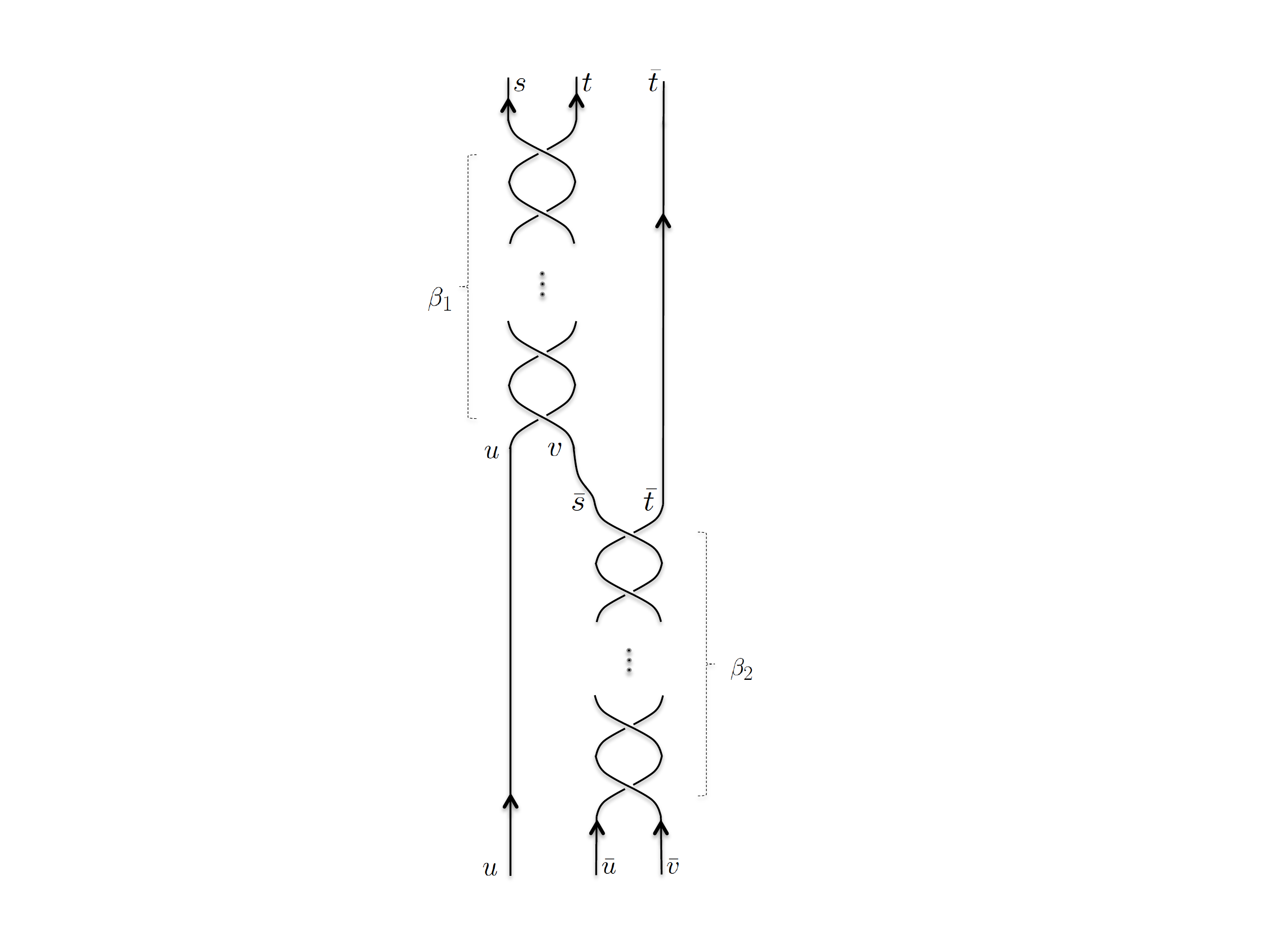}
\caption{This figure depicts a composite braid parametrized by $\beta_1, \beta_2$ (its standard braid closure yields a certain connected sum of two twist links of crossing numbers $\beta_{1}$ and $\beta_2$), and the generators of the fundamental group of its exterior. }
\label{fundBraid1}
\end{figure}

\begin{figure}[h]
\centering
\includegraphics[width=150mm]{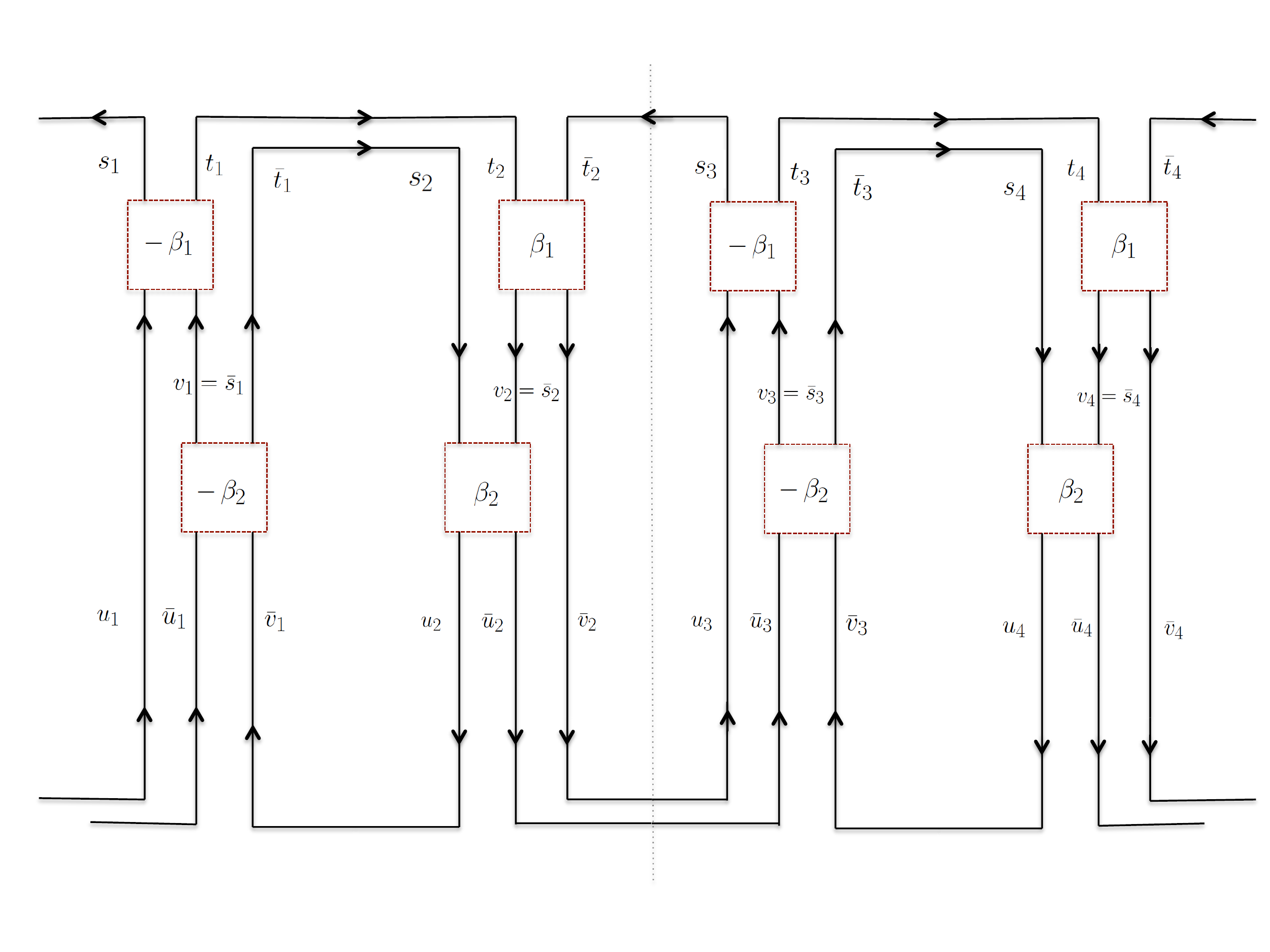}
\caption{We sketch the generators for link $\mathcal{L}_2$ of the composite braid parametrized by $\beta_1, \beta_2$. The open lines at the left and right ends are identified. The dotted line at the center separates the two density matrices. The generalization for $\mathcal{L}_n$ is straightforward and the fundamental group is computed in this section. }
\label{fundBraid2}
\end{figure}

We can similarly derive the Wirtinger presentation of the auxiliary link group for the braid of which closure
yields the sum of two links. Let $\beta_1, \beta_2$ be the number of crossings in a two-strand braid. After similar considerations as above, we find the following Wirtinger presentation for the auxiliary link
$\mathcal{L}_n$. Let $u,\bar{u}_i, \bar{v}_i$ be the generators corresponding to the $i$th segment of the link diagram as indicated in Figure \ref{fundBraid1} and \ref{fundBraid2}.

For definiteness, let begin with the specific case of $\beta_1=2k_1, \beta_2=2k_2$ being both even. 
We find the following expressions for them in terms of the generators $s_i, t_i$. For those associated with the state ket $|\varphi \rangle$, we find from \eqref{w1} and \eqref{w2} that 
\bea
u_i &=& (t_i s_i)^{-k_1} s_i (t_i s_i)^{k_1}, \cr
\bar{u}_i &=& (s^{-1}_{i+1} v_i )^{-k_2} v_i (s^{-1}_{i+1} v_i )^{k_2}, \cr
\bar{v}_i &=& (s^{-1}_{i+1} v_i )^{-k_2+1} v^{-1}_i s^{-1}_{i+1} v_i
(s^{-1}_{i+1} v_i )^{k_2-1},
\eea
where 
$$
v_i = (t_i s_i)^{-k_1+1} s^{-1}_i t_i s_i (t_i s_i)^{k_1 -1}.
$$
For those associated with the state bra $\langle \varphi |$, we find
\bea
u_{i+1} &=& (v_{i+1} s_{i+1})^{k_2 -1} (v_{i+1} s_{i+1} v^{-1}_{i+1} )  (v_{i+1} s_{i+1})^{-k_2 +1},\cr
\bar{u}_{i+1} &=& (v_{i+1} s_{i+1} )^{k_2}
v_{i+1}(v_{i+1} s_{i+1} )^{-k_2}, \cr
\bar{v}_{i+1} &=& (s^{-1}_{i+2} t^{-1}_i)^{k_1} s^{-1}_{i+2} (s^{-1}_{i+2} t^{-1}_i)^{-k_1},
\eea
where $v_{i+1}= (s^{-1}_{i+2} t^{-1}_i )^{k_1-1} s^{-1}_{i+2} t^{-1}_i s_{i+2} 
(s^{-1}_{i+2} t^{-1}_i )^{-k_1+1}$. It is then easy to write down the Wirtinger presentation of the fundamental group of the auxiliary link's exterior which reads
\be
\pi_1 (S^3 - \mathcal{L}_n ) = \langle 
s_1, s_2, \ldots, s_{2n-1}, s_{2n},
t_1, t_3, \ldots, t_{2n-1} | 
\bar{v}_i = u^{-1}_{i+1},  \bar{u}_{2i} \bar{u}_{2i+1} = 1,\,\,\forall i=1,2,\ldots,2n \rangle
\ee
Generally, for $\mathcal{L}_n$, we have $3n$ generators coupled with $3n-1$ relations, since one can check that any one of the relations is implied by all others by virtue of 
\be
\prod^{2n}_{k=1} u_k \bar{u}_k \bar{v}_k =1
\ee
Similar to the case of the two-strand braid,
as a consistency check, let's take the simplest case of $n=1$ in which case the auxiliary link is the split union of three unknots. We have the generators $s_1, s_2, t_1$ satisfying the three relations
\be
\bar{v}_1 = u^{-1}_2, \bar{u}_2 = \bar{u}^{-1}_1, \bar{v}_2 = u^{-1}_1.
\ee
where 
\bea
u_1&=&(t_1 s_1)^{-k_1} s_1 (t_1 s_1)^{k_1}, \cr
u_2&=&(v_2 s_2)^{k_2 - 1} v_2 s_2 v^{-1}_2 (v_2 s_2)^{1-k_2}, \cr
\bar{v}_1 &=& (s^{-1}_2 v_1 )^{1-k_2} v^{-1}_1 s^{-1}_2 v_1 (s^{-1}_2 v_1)^{k_2 -1}, \cr
\bar{v}_2 &=& (s^{-1}_1 t^{-1}_1 )^{k_1} s^{-1}_1 (s^{-1}_1 t^{-1}_1)^{-k_1}\cr
v_1 &=& (t_1 s_1)^{-k_1} t_1 (t_1 s_1)^{k_1} = v^{-1}_2.
\eea
One can proceed to demonstrate that the three relations are identically satisfied. There is no quotient action on the free group of the three generators so the fundamental group is nothing but $\mathbb{Z} \oplus \mathbb{Z} \oplus \mathbb{Z}$ which is expected for the split union of three unknots. In Appendix \ref{sec:AppB},
we complete our discussion by writing down the link group for other parity choices of $\beta_1, \beta_2$.

The Wirtinger presentation is useful for various purposes of further analysis. For example it features in the computation of the Alexander polynomial. We see that both parameters of the composite braid $\beta_1, \beta_2$ are manifest in the Wirtinger presentation. In principle, one can also explore various homomorphisms of the auxiliary link group onto other finite groups such as permutation groups or onto $SL(2,\mathbb{C})$. We leave these issues for future work.


\subsection{On Seifert surfaces of the auxiliary link}

Seifert surfaces furnish an important characterization of links in dfferent manners, in particular in the study of  the factorizability of knots and computation of Alexander polynomials. Recall that a Seifert surface for an oriented link in $S^3$ is a connected compact oriented surface contained in $S^3$ which has the link as its oriented boundary (see for example \cite{Lick}). 

\begin{figure}
\centering
\includegraphics[scale=0.5]{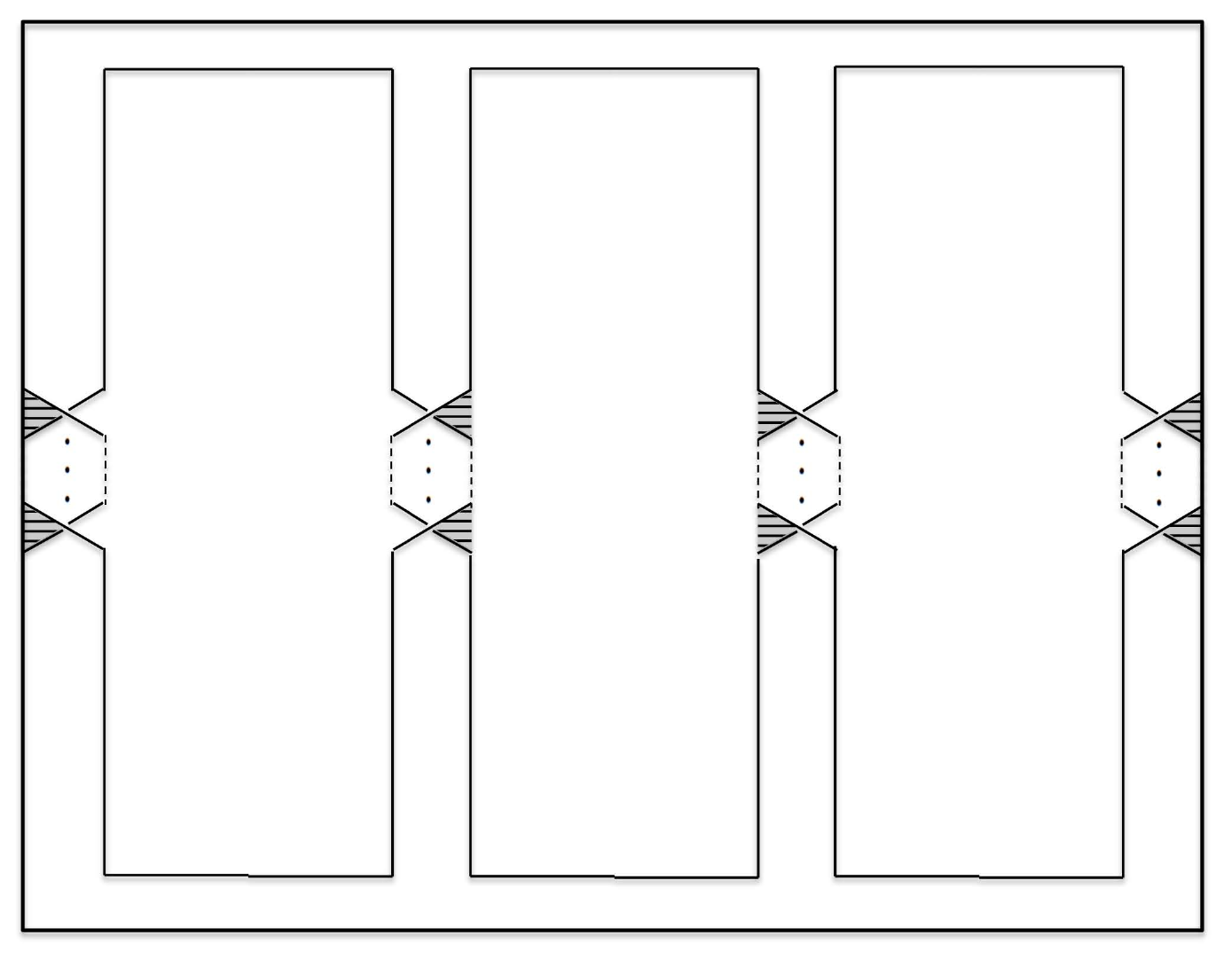}
\caption{Figure sketching a Seifert surface for $\mathcal{L}_n$ of the 2-strand braid for the case of $n=2$. The half-twists ($2n \beta$ of them) join the disjoint circuits at various segments in such a way that the surface is orientable. }
\label{Seifert2strand}
\end{figure}

We shall construct Seifert surfaces for the auxiliary links in a standard algorithmic fashion. For every crossing
in the auxiliary link diagram, in its small neighborhood, we remove the crossing such that its removal is compatible with orientation, yielding a diagram which is the disjoint union of oriented circles. We then 
join these discs or Seifert circles with half-twisted strips at the crossings forming an oriented surface with the auxiliary link as the boundary. The genus of the Seifert surface $F$ is simply 
\be
g(F) = \frac{1}{2} \left(2- m_{components} + m_{crossings} - m_{circles} \right)
\ee
where $m_{components}, m_{circles}$ are the number of link components and Seifert circles contructed by resolving the crossings as shown in Fig. \ref{Seifert2strand}. 
For the auxiliary link associated with $\beta$-crossings of the two-strand braid, we find
$m_{crossings}=2n\beta, m_{circles}=2n$ 
and
\be
m_{components} =\begin{cases}
  2, & \text{for}\,\,\beta \,\,\text{even}\cr    
  2n, & \text{for}\,\,\beta \,\,\text{odd}\\       
\end{cases}
\ee
Thus, the genus of the Seifert surface constructed in the above fashion reads 

\be
g( \mathcal{L}_{\text{2-strand}}) = \begin{cases}
(1+n(\beta -2)), & \text{for}\,\,\beta \,\, \text{even}\cr    
n(\beta -1), & \text{for}\,\,\beta \,\,\text{odd}\\       
\end{cases}
\ee

For the auxiliary link associated with the composite 3-strand braid parametrized by $\beta_1, \beta_2$, we find
$m_{crossings}=2n(\beta_1, \beta_2), m_{circles} = 2n+1$ and 
\be
m_{components} =\begin{cases}
  3, & \text{for}\,\,\beta_1, \beta_2 \,\,\text{odd}\cr    
  2n+1, & \text{for all other parity combination of} \,\,\beta_1, \beta_2\\       
\end{cases}
\ee
Thus, the genus of the Seifert surface in this case reads 

\be
g( \mathcal{L}_{\text{3-strand}}) = \begin{cases}
n(\beta_1 + \beta_2)-(n+1), & \text{for}\,\,\beta_1, \beta_2 \,\, \text{odd}\cr    
n(\beta_1 + \beta_2 -2) , & \text{for all other parity combination of} \,\,\beta_1, \beta_2\\       
\end{cases}
\ee

\begin{figure}
\centering
\includegraphics[scale=0.5]{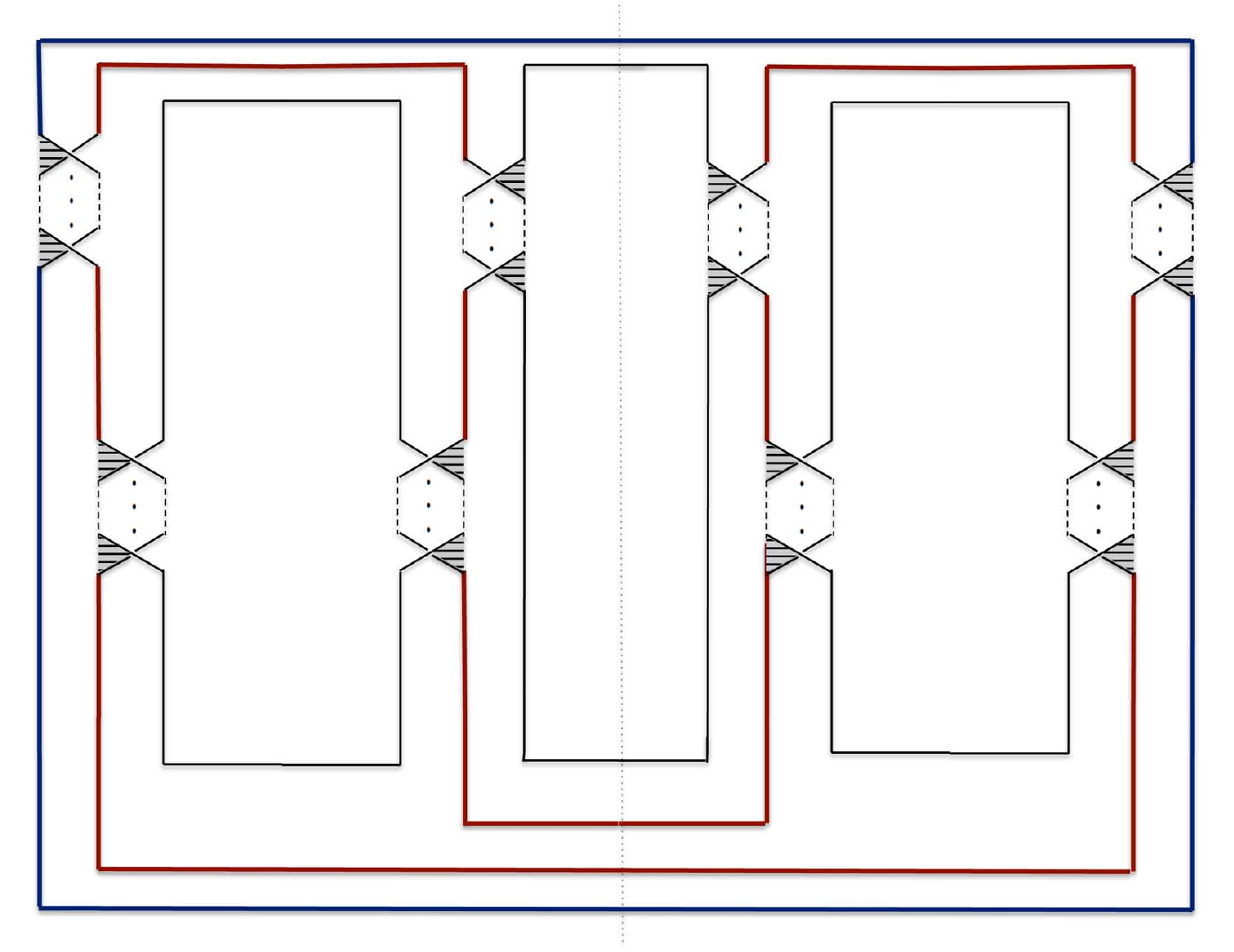}
\caption{Figure sketching a Seifert surface for $\mathcal{L}_n$ of the 3-strand braid for the case of $n=2$. The half-twists join the disjoint circuits at various segments such that there are $2n+1$ Seifert circles in their absence. We find that the genus of this Seifert surface scales linearly with $\beta_{1,2}$. }
\label{Seifert3strand}
\end{figure}

These are upper bounds for the genus of the respective auxiliary links. Recall that it is an important result in knot theory that for any two knots $K_1$ and $K_2$, the genus is additive, i.e.
\be
g(K_1 + K_2) = g(K_1) + g(K_2).
\ee
In this case, for the composite braid we see that the number of crossings in each braid is additive in their appearance in the genus of the Seifert surface which is an upper bound to the genus of the underlying auxiliary link. It would be interesting to study the latter property further, noting that the various observations made in this section are sensitive to the choice of the region $\mathcal{R}_B$. Seifert matrices can be constructed similarly and they lead to the Alexander polynomials of the auxiliary link which are related in computing Wilson loop VEVs in Chern-Simons theory with $U(1,1)$ gauge group \cite{Roz}.


\section{Discussion}
\label{sec:Discussion}

We have explored various ways by which topological entanglement entropy in Chern-Simons theory probes 
the braiding of quasi-particles, based on a bipartition of the system specified by a choice of the 
region $R_{\mathcal{B}}$. By gluing together punctured discs on distinct copies of three-balls, we map 
the problem of computing the Renyi entropy to calculating the trace of an auxiliary link in $S^3$. 
The auxiliary link is defined for each choice of the region $R_{\mathcal{B}}$, and is endowed with the symmetry property of being fully amphichiral. We compute the entanglement
measures for a few simple cases: the 2-strand braid with arbitrary number of crossings and the connected sum
of two Hopf links which admits a 3-strand braid representation. The computation essentially reduces to that of the Jones polynomial of the auxiliary link, and relies on us being able to express it as a function of the power index of the density matrix. 
Getting such a computation performed allows us to study how the entanglement measures distinguish between
different braided configurations by being distinct functions of the Chern-Simons level $k$.

Apart from calculating the Chern-Simons VEV, we examined a couple of elementary topological 
aspects of the auxiliary link. For the cases considered in this work, we computed the fundamental group of the auxiliary link's exterior, and thus furnished an exact
description of how it is sensitive to the braiding parameters. The genus of the Seifert surface constructed by algorithmically
resolving crossings is linear in the crossing number which is additive under the connected sum. The fundamental group and the Seifert matrices associated with the Seifert surfaces would be useful in determining other properties of the auxiliary link such as its Alexander polynomial which would feature in the Chern-Simons theory with a suitable super-Lie group such as $U(1,1)$ \cite{Roz}. 

Future directions naturally include exploring similar ideas for Chern-Simons theories with other gauge groups and studying other entanglement measures. For example, it would be natural to assume that some form of charged entanglement entropy
can be defined in the context of refined Chern-Simons theory, and we are then poised to ask how topological entanglement entropy may contain knot homological information \cite{Mina}. Previously, it was found that braided tensor categorical descriptions of TQFTs appear to be a natural language for studies of topological entanglement entropy. It would be nice to understand how to formulate our various results in such a framework, by for example rephrasing the VEV of the auxiliary link in terms of fusion and R-matrices, etc. This would in principle lead to a CFT description of our results. We hope to report on this soon \cite{Tan}.

Another natural future direction would be to consider the theory on other 3-manifolds other than spheres. In \cite{Fradkin1}, the entanglement measures were also computed for $T^2$ for various interfaces without the inclusion of quasi-particles, essentially invoking the surgery formula $Z(M) = Z(M_1) Z(M_2)/Z(S^3)^n$ where $M$ is a 3-manifold that is the connected
sum of $M_1$ and $M_2$ joined along $n$ two-spheres (see \cite{Ryu1} for the corresponding edge theoretic approach). It would be interesting to consider the presence of Wilson lines joining quasi-particles for the theory on higher-genus Riemann surfaces. Broadly speaking, we hope that our work has furnished another starting point for further explorations of the intricate relations between topological entanglement entropy and the theory of braids, knots and links.

\section*{Acknowledgments}
I am very grateful to Neal Snyderman, Sergey Cherkis, Ori Ganor and Petr Ho\v{r}ava for stimulating discussions and various advice on related topics. The final stages of this work were completed during a short summer visit to Enrico Fermi Institute at University of Chicago. I would like to express my gratitude for their hospitality, especially to Savdeep Sethi for being immensely inspirational. Finally, I acknowledge support from a research fellowship given by the School of Physical and Mathematical Sciences, Nanyang Technological University of Singapore.


\appendix

\section{On the auxiliary link for the 2-strand braid with two crossings (Hopf link)}

In this Appendix, we present a short note on another derivation of $\mathcal{L}_n$ for the 2-strand braid with two crossings as an independent check of the formula \ref{twostrandRenyi} presented in the main text. The derivation relies on resolving crossings sequentially from one end to the other until we end up with a 3 component link which we identify to be $\bar{6}^3_3$ in Rolfsen's table (see for example \cite{knotatlas}. Taking $n=2$ yields the $8^4_3$ link of which tabulated Jones polynomial is checked to agree with the one obtained here. The details are as follows, with sketches of $\bar{6}^3_3$ and $8^4_3$ in Fig. \ref{fig:Appendix1} below. 
Using the following skein relation we find
\bea
\mathcal{L}_{n} &=& \bar{A}^2 H_{2n} + A^2 (1-A^4)^2 H_{2n-2} \sum_{j=0}^{n-4}r^j + 
\bar{A}^2 (1-A^4)H_{2n-1} \sum_{j=0}^{n-3} r^j  \cr
&&+ (1-A^4)^2| 1-A^4|^{2n-6} \left(  (1- \bar{A}^4) \langle \bar{6}^3_3 \rangle + A^2 H_4 \right) \nonumber \\
&=& (1-A^4)^2 (1-A^4)^{n-3} (1-\bar{A}^4)^{n-3} (1- \bar{A}^4) \langle \bar{6}^3_3 \rangle 
+ \bar{A}^{2} H_{2n} \cr
&&+ \left( \frac{1-r^{n-2}}{1-r} \right) \left(  H_{2n-2} A^2 (1-A^4)^2 + \bar{A}^2 (1-A^4)H_{2n-1} \right)
\eea
where $r = \frac{|1-A^4|^2}{(A^4 + \bar{A}^4)^2}$, $H_k = (-A^4 - \bar{A}^4 )^{k-1}$ is the bracket polynomial for a $k$-component Hopf link and $\langle \bar{6}^3_3 \rangle $ is the bracket polynomial for the $\bar{6}^3_3$ link.  Upon simplifying, we obtain 
\be
\label{twostrandOnecrossing}
\mathcal{L}_{n} = -\frac{ (A^4 + A^{-4})^{2n} }{A^2 + A^{-2}} + (-1)^{n+1} (A^2 - A^{-2})^{2n}
\frac{  1+A^4+A^{-4}          }{A^2 + A^{-2}} 
\ee
valid for all $n\geq 1$. Some consistency check: for $n=2$, this link is $8^4_3$ in Rolfsen's table, and 
it is easy to check that \eqref{twostrandOnecrossing} reduces to  
\be
L_4 = -A^{-14} (1 + A^8 + 2 (A^{12} + A^{16}) + A^{20} + A^{28} )
\ee
which is the polynomial as tabulated in \cite{knotatlas}. 

\begin{figure}[h]
    \centering
    \includegraphics[width=0.35\textwidth]{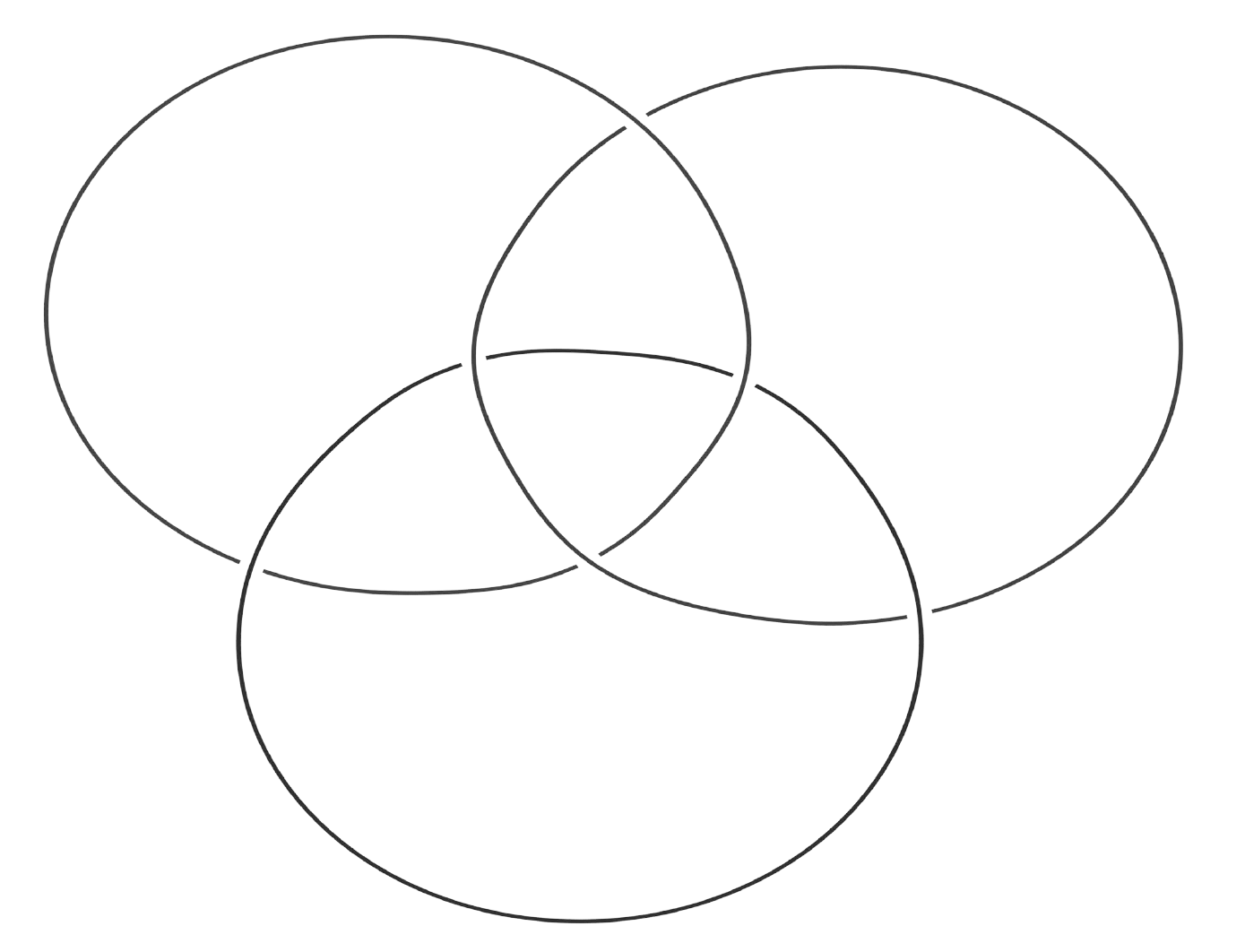}
    \includegraphics[width=0.3\textwidth]{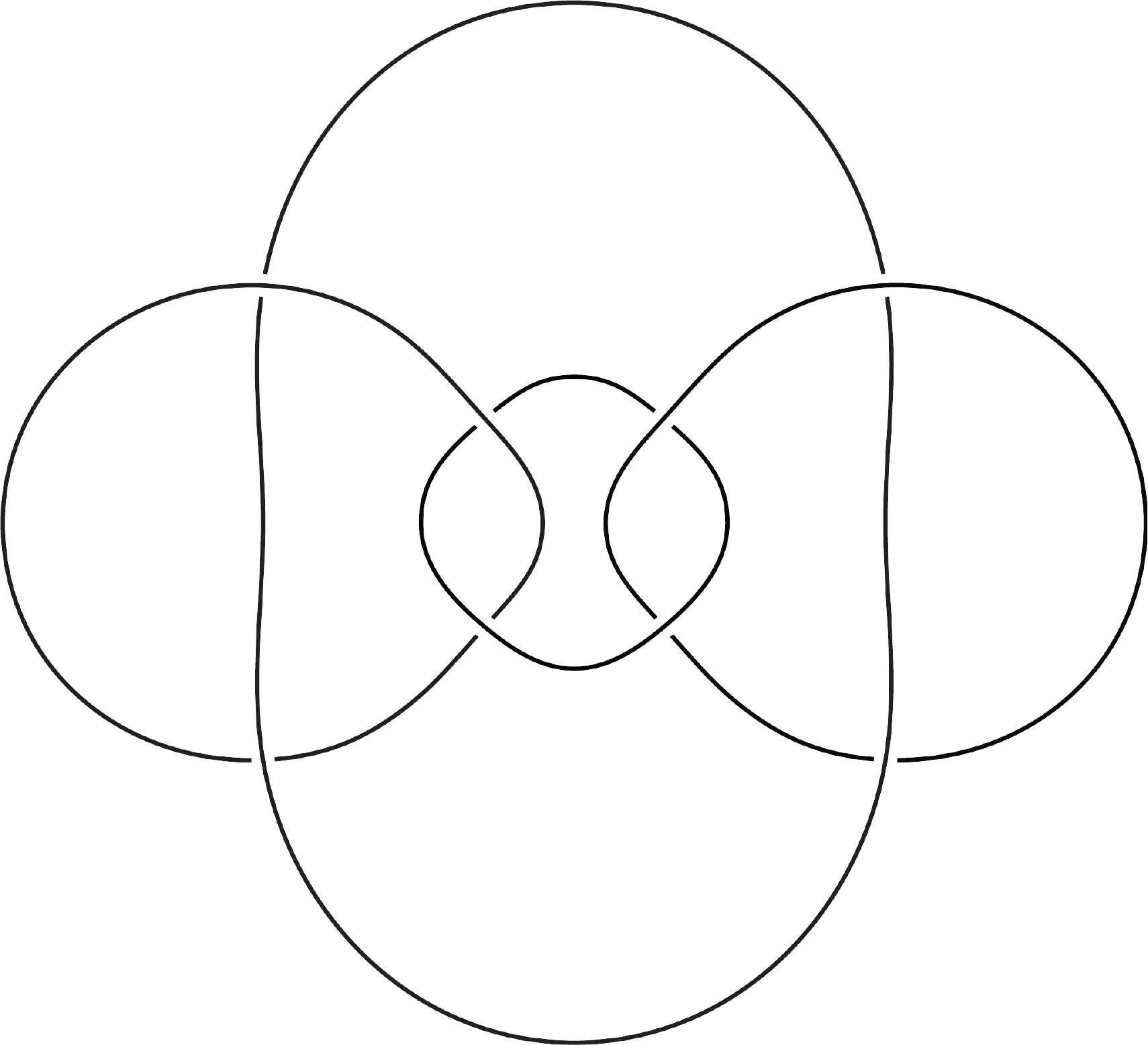}
    \caption{Sketches of the $\bar{6}^3_3$ (left) and $8^4_3$ (right) links. The latter is equivalent to $\mathcal{L}_2$ for the 2-strand braid presentation of the Hopf link.}
    \label{fig:Appendix1}
\end{figure}


\section{More about the auxiliary link group of the 3-strand braid}
\label{sec:AppB}

In this Appendix, we complete our discussion of the fundamental group of the auxiliary link's exterior in Section \ref{sec:group} by considering other parity choices of $\beta_1, \beta_2$ in the 3-strand braid. 
We express them using the same set of $3n$ generators, and the $3n-1$ relations in similar form, i.e
\be
\pi_1 (S^3 - \mathcal{L}_n ) = \langle 
s_1, s_2, \ldots, s_{2n-1}, s_{2n},
t_1, t_3, \ldots, t_{2n-1} | 
\bar{v}_i = u^{-1}_{i+1},  \bar{u}_{2i} \bar{u}_{2i+1} = 1,\,\,\forall i=1,2,\ldots,2n \rangle,
\ee
with any one of the relations being implied by all others by virtue of 
\be
\prod^{2n}_{k=1} u_k \bar{u}_k \bar{v}_k =1.
\ee
For other parity choices of $\beta_1, \beta_2$, each comes
with different dependence of $u_i, \bar{u}_i, v_i, \bar{v}_i$ on the generators. 
\\[2ex]
\noindent
(i)\underline{\textbf{Even $\beta_1$ and odd $\beta_2$}} \\[1ex]
For variables associated with the state ket $|\varphi \rangle$, we find
\bea
u_i &=& (t_i s_i)^{-k_1} s_i (t_i s_i)^{k_1}, \cr
\bar{v}_i &=& (s^{-1}_{i+1} v_i )^{-k_2} v_i (s^{-1}_{i+1} v_i )^{k_2}, \cr
\bar{u}_i &=& (s^{-1}_{i+1} v_i )^{-k_2} (v^{-1}_i s^{-1}_{i+1} v_i
(s^{-1}_{i+1} v_i )^{k_2},
\eea
where 
$$
v_i = (t_i s_i)^{-k_1+1} s^{-1}_i t_i s_i (t_i s_i)^{k_1 -1}.
$$
For those associated with the state bra $\langle \varphi |$, we find
\bea
\bar{u}_{i+1} &=& (v_{i+1} s_{i+1})^{k_2 } (v_{i+1} s_{i+1} v^{-1}_{i+1} )  (v_{i+1} s_{i+1})^{-k_2 },\cr
u_{i+1} &=& (v_{i+1} s_{i+1} )^{k_2}
v_{i+1}(v_{i+1} s_{i+1} )^{-k_2}, \cr
\bar{v}_{i+1} &=& (s^{-1}_{i+2} t^{-1}_i)^{k_1} s^{-1}_{i+2} (s^{-1}_{i+2} t^{-1}_i)^{-k_1},
\eea
where $v_{i+1}= (s^{-1}_{i+2} t^{-1}_i )^{k_1-1} s^{-1}_{i+2} t^{-1}_i s_{i+2} 
(s^{-1}_{i+2} t^{-1}_i )^{-k_1+1}$. 
\\[2ex]
\noindent
(ii)\underline{\textbf{Odd $\beta_1$ and even $\beta_2$}} \\[1ex]
For variables associated with the state ket $|\varphi \rangle$, we find
\bea
u_i &=& (t_i s_i)^{-k_1} s^{-1}_i t_i s_i (t_i s_i)^{k_1}, \cr
\bar{u}_i &=& (s^{-1}_{i+1} v_i )^{-k_2} v_i (s^{-1}_{i+1} v_i )^{k_2}, \cr
\bar{v}_i &=& (s^{-1}_{i+1} v_i )^{-k_2+1} (v^{-1}_i s^{-1}_{i+1} v_i
(s^{-1}_{i+1} v_i )^{k_2},
\eea
where 
$$
v_i = (t_i s_i)^{-k_1} s_i  (t_i s_i)^{k_1}.
$$
For those associated with the state bra $\langle \varphi |$, we find
\bea
u_{i+1} &=& (v_{i+1} s_{i+1})^{k_2 -1 } (v_{i+1} s_{i+1} v^{-1}_{i+1} )  (v_{i+1} s_{i+1})^{1-k_2 },\cr
\bar{u}_{i+1} &=& (v_{i+1} s_{i+1} )^{k_2}
v_{i+1}(v_{i+1} s_{i+1} )^{-k_2}, \cr
\bar{v}_{i+1} &=& (s^{-1}_{i+2} t^{-1}_i)^{k_1} s^{-1}_{i+2} t^{-1}_i s_{i+2} (s^{-1}_{i+2} t^{-1}_i)^{-k_1},
\eea
where $v_{i+1}= (s^{-1}_{i+2} t^{-1}_i )^{k_1} s^{-1}_{i+2} (s^{-1}_{i+2} t^{-1}_i )^{-k_1}$. 
\\[2ex]
\noindent
(iii)\underline{\textbf{Both $\beta_1, \beta_2$ odd}} \\[1ex]
For variables associated with the state ket $|\varphi \rangle$, we find
\bea
u_i &=& (t_i s_i)^{-k_1} s^{-1}_i t_i s_i (t_i s_i)^{k_1}, \cr
\bar{v}_i &=& (s^{-1}_{i+1} v_i )^{-k_2} v_i (s^{-1}_{i+1} v_i )^{k_2}, \cr
\bar{u}_i &=& (s^{-1}_{i+1} v_i )^{-k_2} (v^{-1}_i s^{-1}_{i+1} v_i
(s^{-1}_{i+1} v_i )^{k_2},
\eea
where 
$$
v_i = (t_i s_i)^{-k_1} s_i (t_i s_i)^{k_1}.
$$
For those associated with the state bra $\langle \varphi |$, we find
\bea
\bar{u}_{i+1} &=& (v_{i+1} s_{i+1})^{k_2 } (v_{i+1} s_{i+1} v^{-1}_{i+1} )  (v_{i+1} s_{i+1})^{-k_2 },\cr
u_{i+1} &=& (v_{i+1} s_{i+1} )^{k_2}
v_{i+1}(v_{i+1} s_{i+1} )^{-k_2}, \cr
\bar{v}_{i+1} &=& (s^{-1}_{i+2} t^{-1}_i)^{k_1} s^{-1}_{i+2}t^{-1}_i s_{i+2} (s^{-1}_{i+2} t^{-1}_i)^{-k_1},
\eea
where $v_{i+1}= (s^{-1}_{i+2} t^{-1}_i )^{k_1} s^{-1}_{i+2}  
(s^{-1}_{i+2} t^{-1}_i )^{-k_1}$.

\end{document}